\documentclass[aps,prb,superscriptaddress,twocolumn,preprintnumbers,amsmath,amssymb]{revtex4}
\usepackage{color}
\usepackage{array}
\usepackage{amsmath}
\usepackage{amssymb}
\usepackage{epsfig}
\usepackage{graphicx}
\usepackage{bbm}

\newcommand{\Ref}[1]{Ref.~\onlinecite{#1}}
\newcommand{\bs}[1]{{\boldsymbol#1}}
\newcommand{\bss}{{\boldsymbol{\sigma}}}
\newcommand{\ie}{{\emph{i.e.~}}}
\makeatletter
\newcommand{\rmnum}[1]{\romannumeral #1}
\newcommand{\Rmnum}[1]{\expandafter\@slowromancap\romannumeral #1@}
\makeatother
\newcommand{\imth}{\hspace{1pt}\mathrm{i}\hspace{1pt}}
\newcommand{\alert}[1]{{\color{red}{#1}}}
\newcommand{\eg}{{\emph{e.g.~}}}
\newcommand{\etc}{{\emph{etc.~}}}

\begin{document}
\title{Spin liquids on a honeycomb lattice: Projective Symmetry Group study of Schwinger fermion mean-field theory}
\author{Yuan-Ming Lu}
\author{Ying Ran}
\affiliation{Department of Physics, Boston College, Chestnut Hill, MA 02467}
\date{\today}

\begin{abstract}
Spin liquids are novel states of matter with fractionalized
excitations. A recent numerical study of Hubbard model on a
honeycomb lattice\cite{Meng2010} indicates that a gapped spin liquid
phase exists close to the Mott transition. Using Projective Symmetry
Group, we classify all the possible spin liquid states by Schwinger
fermion mean-field approach. We find there is only one fully gapped
spin liquid candidate state: ``Sublattice Pairing State'' that can
be realized up to the 3rd neighbor mean-field amplitudes, and is in
the neighborhood of the Mott transition. We propose this state as
the spin liquid phase discovered in the numerical work. To
understand whether SPS can be realized in the Hubbard model, we
study the mean-field phase diagram in the $J_1-J_2$ spin-1/2 model
and find an $s$-wave pairing state. We argue that $s$-wave pairing
state is not a stable phase and the true ground state may be SPS. A
scenario of a continuous phase transition from SPS to the semimetal
phase is proposed. This work also provides guideline for future
variational studies of Gutzwiller projected wavefunctions.
\end{abstract}

\maketitle

\section{Introduction}

Traditional Landau's theory\cite{landau_book,landau_ginzburg} points
out that states of matter can be classified by their symmetry. And
the low energy excitations can be understood by either bosonic modes
or fermionic quasiparticles, which carry integer multiples of the
quantum numbers of the fundamental degrees of freedom. Fractional
quantum Hall liquids (FQHLs) provide a striking counterexample of
the Laudau's paradigm: different FQHLs all have the same symmetry,
yet they are very different since a quantum phase transition is
required to go from one liquid to another. To understand their
differences, one has to go beyond Laudau's paradigm and the concept
of topological order was
introduced\cite{PhysRevB.41.9377,wen_topological_order}. The
quasiparticle excitations in FQHLs carry only a fraction of the
fundamental electric charge. Meanwhile these fractionalized
quasiparticles obey neither bosonic nor fermionic statistics and are
dubbed anyons consequently.

Can strong interactions lead to similar novel states of matter in
the absence of magnetic field? After the original proposal of
Anderson\cite{Anderson:1987gf}, intensive theoretical studies have revealed that
spin systems can realize such novel phases of matter: spin liquids(SL).
And a few experimental systems have been identified to be likely in
spin liquid phases\cite{PhysRevLett.91.107001,helton-2007-98,PhysRevLett.99.137207}. A spin liquid is often
defined to be a quantum phase of spin-1/2 per unit cell that does
not break translation symmetry. These liquid phases of spins are also
distinct from one another by their topological order. Although a
rigorous theorem is lacking because we are still unable to classify
all possible topological order, it is generally believed that the
excitation of a topological ordered phase is
fractionalized\cite{PhysRevLett.96.060601}.

Although theoretical studies have shown that spin liquid ground
states exist for artificial model
Hamiltonians\cite{PhysRevB.37.3774,PhysRevLett.66.1773,PhysRevLett.61.2376,PhysRevLett.86.1881,Kitaev20062,PhysRevLett.90.016803},
it remains unclear whether a simple or experimentally realizable
Hamiltonian can host such novel states. Recently a remarkable
quantum Monte Carlo simulation of Hubbard model on a honeycomb
lattice\cite{Meng2010} indicates that a gapped spin disordered
ground state exists in the neighborhood of the Mott transition.
Although a honeycomb lattice has two spin-1/2 per unit cell, it is
impossible to have a band insulator phase without breaking lattice
symmetry. Therefore this spin disordered phase should be
topologically ordered and have fractionalized excitations. We will
still term it a spin liquid.

What is the nature of this spin liquid phase? In this paper we try
to propose the candidate states by Schwinger-fermion (or
slave-boson) mean-field
approach\cite{2986272,PhysRevB.37.580,PhysRevB.37.3774,3246071,PhysRevB.38.745,PhysRevLett.76.503,9015906},
following the techniques developed in \Ref{PhysRevB.65.165113}. Our
results can be summarized as follows. We search for the fully gapped
spin liquids which lead us to focus on the $Z_2$ mean-field states.
We first use Projective Symmetry Group
(PSG)\cite{PhysRevB.65.165113} to classify all 128 possible $Z_2$
mean-field states that preserve the full lattice symmetry as well as
time-reversal symmetry. Notice the spin liquid phase in the
numerical work seems to be connected to the semi-metal phase by a
second-order phase transition, which suggests this state to be in
the neighborhood of a uniform Resonating-Valence-Bond (u-RVB) state.
So we classify all the 24 possible $Z_2$ states around the u-RVB
states. Among these 24 states, we find only 4 states can have a full
energy gap in the spinon spectrum, while other 20 states have
symmetry protected gapless spinon excitations. We find that up to
3rd neighbor mean-field amplitudes, only one of the four fully
gapped $Z_2$ state can be realized, and we term it as Sublattice
Pairing State (SPS). We propose this state to be the spin liquid
state discovered in the numerical study. We also study the
mean-field phase diagram of the $J_1-J_2$ antiferromagnetic spin-1/2
model on a honeycomb lattice to understand whether SPS can be more
favorable than the u-RVB state while both states are in the
neighborhood of the Mott transition. We find when $J_2> 0.85 J_1$ a
spinon gap opens up by $s$-wave pairing on top of the u-RVB state.
This $s$-wave pairing state is not a stable phase and is an artifact
of the mean-field study where gauge dynamics are ignored. On the
other hand, the proposed SPS $Z_2$ state is continuously connected
to the $s$-wave pairing state by making the pairing phase sublattice
dependent. This suggests the ultimate fate of $s$-wave pairing state
may be SPS. We propose that a more careful projected wavefunction
study, which includes the gauge fluctuations, may be able to find
SPS $Z_2$ state as the ground state in the $J_1-J_2$ model. The
possible continuous phase transitions from SPS into semi-metal phase
are discussed.

\section{Schwinger-fermion approach and PSG}\label{SF_PSG}

In Schwinger-fermion approach, a spin-1/2 operator at
site $i$ is represented by:
\begin{align}
\vec
S_i=\frac{1}{2}f_{i\alpha}^{\dagger}\vec\sigma_{\alpha\beta}f_{i\beta}.\label{eq:schwinger-fermion}
\end{align}
A Heisenberg spin Hamiltonian $H=\sum_{<ij>} J_{ij} \vec S_i\cdot
\vec S_j$ is represented as
$H=\sum_{<ij>}-\frac{1}{2}J_{ij}\big(f_{i\alpha}^{\dagger}f_{j\alpha}f_{j\beta}^{\dagger}f_{j\beta}+\frac{1}{2}f_{i\alpha}^{\dagger}f_{i\alpha}f_{j\beta}^{\dagger}f_{j\beta}\big)$.
Because this representation enlarges the Hilbert space, states need
to be constrained in the physical Hilbert space, i.e., one
$f$-fermion per site:
\begin{align}
 f_{i\alpha}^{\dagger}f_{i\alpha}&=1,&f_{i\alpha}f_{i\beta}\epsilon_{\alpha\beta}=0.\label{eq:constraint}
\end{align}
Introducing mean-field parameters
$\eta_{ij}\epsilon_{\alpha\beta}=-2\langle
f_{i\alpha}f_{j\beta}\rangle$,
$\chi_{ij}\delta_{\alpha\beta}=2\langle
f_{i\alpha}^{\dagger}f_{j\beta}\rangle$, where
$\epsilon_{\alpha\beta}$ is fully antisymmetric tensor, after
Hubbard-Stratonovich transformation, the Lagrangian of the spin
system can be written as\cite{PhysRevB.65.165113}
\begin{align}
L=&\sum_i\psi_{i}^{\dagger}\partial_{\tau}\psi_{i}+\sum_{<ij>}\frac{3}{8}J_{ij}\big[\frac{1}{2}\mbox{Tr}(U^{\dagger}_{ij}U_{ij})\notag\\
&-(\psi^{\dagger}_i U_{ij}\psi_{j}+h.c.)\big]+\sum_i a_0^l(i)
\psi_i^{\dagger}\tau^l\psi_i\label{eq:action}
\end{align}
where two-component fermion notation
$\psi_i=(f_{i,\uparrow},f_{i,\downarrow}^{\dagger})^T$ is introduced
for reasons that will be explained shortly. $U_{ij}$ is a matrix of
mean-field amplitudes:
\begin{align}
 U_{ij}=\begin{pmatrix}\chi_{ij}^{\dagger}&\eta_{ij}\\\eta_{ij}^{\dagger}&-\chi_{ij}\end{pmatrix}.
\end{align}
$a_0^{l}(i)$ are the local Lagrangian multipliers that enforces the
constraints Eq.(\ref{eq:constraint}).

In terms of $\psi$, Schwinger-fermion representation has an explicit
$SU(2)$ gauge redundancy: a transformation $\psi_i\rightarrow
W_i\psi_i$, $U_{ij}\rightarrow W_i U_{ij}W_j^{\dagger}$, $W_i\in
SU(2)$ leaves the action invariant. This redundancy is originated
from representation Eq.(\ref{eq:schwinger-fermion}): this local
$SU(2)$ transformation leaves the spin operators
invariant\cite{PhysRevB.38.745} and thus does not change physical Hilbert
space.

One can try to solve Eq.(\ref{eq:action}) by mean-field (or
saddle-point) approximation. At mean-field level, $U_{ij}$ and
$a_0^l$ are treated as complex numbers, and $a_0^l$ must be chosen
such that constraints Eq.(\ref{eq:constraint}) are satisfied at the
mean field level: $\langle\psi_i^{\dagger}\tau^l \psi_i\rangle=0$.
The mean-field ansatz can be written as:
\begin{align}
 H_{MF}=-\sum_{<ij>}\psi^{\dagger}_iu_{ij}\psi_{j}+\sum_{i}\psi_i^{\dagger}a_0^l\tau^l\psi_i.\label{eq:mf}
\end{align}
where $u_{ij}=\frac{3}{8}J_{ij}U_{ij}$.   A local $SU(2)$ gauge
transformation modify $u_{ij}\rightarrow W_iu_{ij}W_j^{\dagger}$ but
does not change the physical spin state described by the mean-field
ansatz. By construction the mean-field amplitudes do not break spin
rotation symmetry, and the mean field solutions describe spin liquid
states if translational symmetry is preserved. Different
$\{u_{ij}\}$ ansatz may be in different spin liquid phases. The
mathematical language to classify different spin liquid phases is
PSG\cite{PhysRevB.65.165113}.

\begin{figure}
 \includegraphics[width=0.28\textwidth]{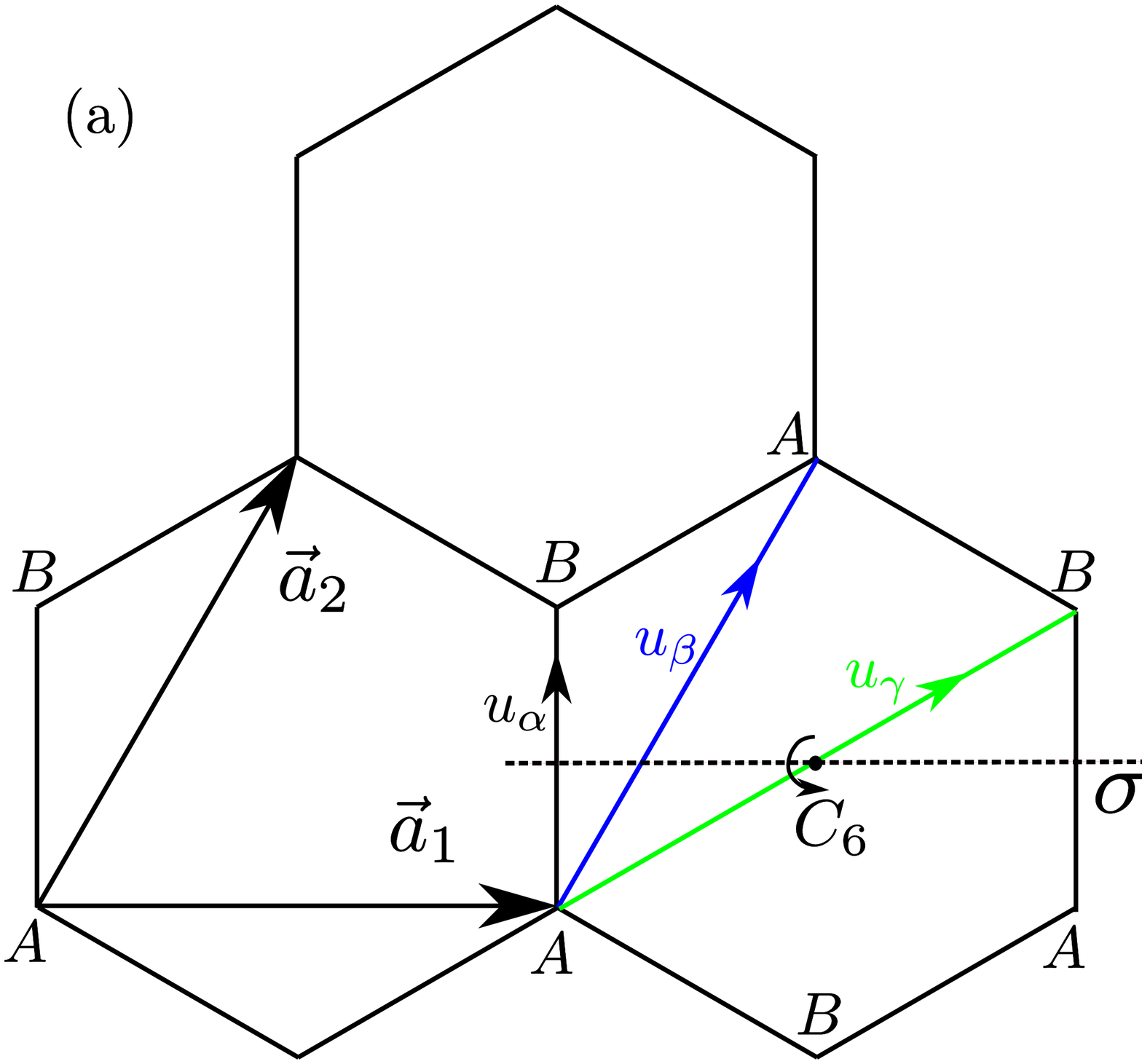}\;\;\;\;\;\includegraphics[width=0.18\textwidth]{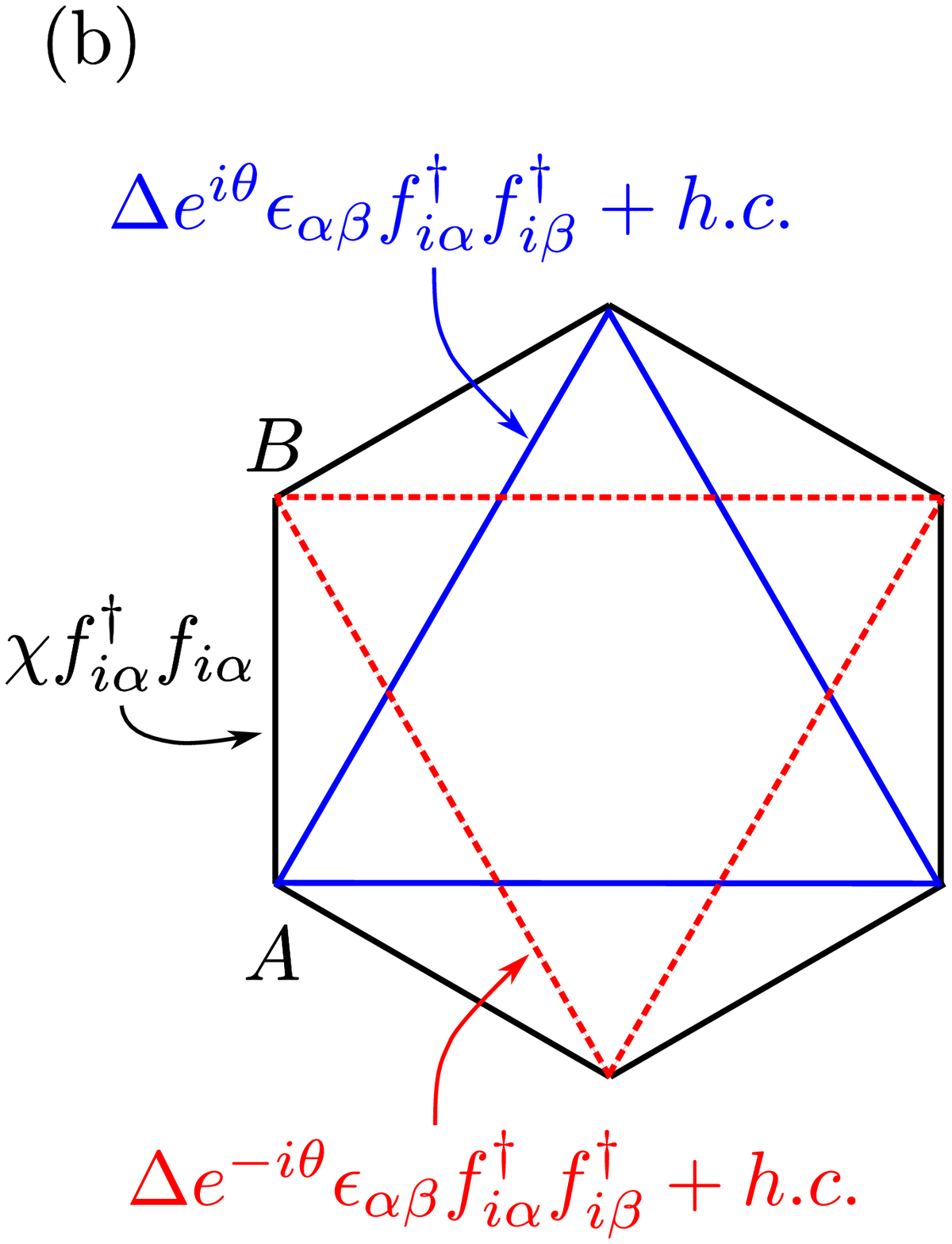}
\caption{(color online) (a) Honeycomb lattice and generators of
symmetry group. (b) SPS mean-field ansatz in terms of $f$-fermion.
$t,\Delta,\theta$ are real and $\theta\neq 0,\pi,\pm\pi/2$. The pairing phases for A sublattice (blue solid line) and B sublattice (red dashed line) are opposite.}
\label{fig:honeycomb}
\end{figure}

PSG is the manifestation of topological order in the
Schwinger-fermion representation: spin liquid states described by
different PSG's are different phases. It is defined as the
collection of all combinations of symmetry group and $SU(2)$ gauge
transformations that leave $\{u_{ij}\}$ invariant (as $a_0^l$ are
determined self-consistently by $\{u_{ij}\}$, these transformations
also leave $a_0^l$ invariant).  The invariance of a mean-field
ansatz $\{u_{ij}\}$ under an element of PSG $G_U U$ can be written
as
\begin{align}
G_U U(\{ u_{ij}\})&=\{u_{ij}\},\notag\\
U(\{u_{ij}\})&\equiv \{\tilde u_{ij}=u_{U^{-1}(i),U^{-1}(j)}\},\notag\\
G_U(\{ u_{ij}\})&\equiv \{\tilde u_{ij}=G_U(i)u_{ij}G_U(j)^{\dagger}\}, \notag\\
&G_U(i)\in SU(2).
\end{align}
Here $U\in SG$ is an element of symmetry group (SG) of the spin
liquid state. SG on a honeycomb lattice is generated by time
reversal $\bs T$, reflection $\bss$, $\pi/3$ rotation $C_6$ and
translations $T_1,T_2$ as illustrated in FIG. \ref{fig:honeycomb}
(see also appendix \ref{app:PSG General conditions}).  $G_U$ is the gauge transformation
associated with $U$ such that $G_U U$ leaves $\{ u_{ij}\}$
invariant.

There is an important subgroup of PSG, Invariant Gauge Group (IGG),
which is composed of all the pure gauge transformations in PSG:
$IGG\equiv\{\{W_i\}|W_iu_{ij}W_j^{\dagger}=u_{ij}, W_i\in SU(2)\}$.
One can always choose a gauge in which the elements in IGG is
site-independent. In this gauge, IGG can be global $Z_2$
transformations: $\{W_i=\tau^0, W_i=-\tau^0\}$, the global $U(1)$
transformations: $\{W_i=e^{i\theta\tau^3},\theta\in[0,2\pi]\}$, or
the global $SU(2)$ transformations: $\{W_i=e^{i\theta\hat n\cdot
\vec\tau},\theta\in[0,2\pi],\hat n\in S^2\}$, and we dub them $Z_2$,
$U(1)$ and $SU(2)$ state respectively.

The importance of IGG is that it controls the low-energy gauge
fluctuations. Beyond mean-field level, fluctuations of $U_{ij}$ and
$a_0^l$ need to be considered and the mean-field state may or may
not be stable. The low-energy effective theory is described by
fermionic spinon band structure coupled with a dynamical gauge field
of IGG. For example, $Z_2$ state with gapped spinon dispersion can
be a stable phase because the low-energy $Z_2$ dynamical gauge field
can be in the deconfined phase\cite{wegner_Z2,RevModPhys.51.659}. But for a $U(1)$ state with
gapped spinon dispersion, the $U(1)$ gauge fluctuations would
generally drive the system into confinement due to monopole
proliferation\cite{Polyakov}, and the mean-field state would be
unstable. And an $SU(2)$ state with gapped spinon dispersion should
also be in the confined phase because there is no known IR stable
fixed point of pure $SU(2)$ gauge theory in 2+1 dimension. Because
the purpose of this paper is to search for stable spin liquid phases
that has a Schwinger fermion mean-field description, we will focus
on $Z_2$ states.

If $G_UU\in PSG$ and $g\in IGG$, by definition we have $gG_UU\in
PSG$. This means that the mapping $h: PSG\rightarrow SG: f(G_UU)=U$
is a many-to-one mapping. In fact it is easy to show that mapping
$h$ induces group homomorphism\cite{PhysRevB.65.165113}:
\begin{align}PSG/IGG=SG.\label{eq:PSG}\end{align} Mathematically
$PSG$ is an extension of $SG$ by $IGG$.

Our definition of PSG requires a mean-field ansatz $\{u_{ij}\}$.
With Eq.(\ref{eq:PSG}), one can define algebraic-PSG which does not
require ansatz $\{u_{ij}\}$. An algebraic-PSG is simply defined as a
group satisfying Eq.(\ref{eq:PSG}). Obviously a PSG (realizable by
an ansatz) must be an algebraic-PSG, but the reverse may not be
true, because sometimes an algebraic-PSG cannot be realized by any
mean-field ansatz.

To classifying all possible $Z_2$ Schwinger-fermion mean-field
states, we need to find all possible $PSG$ group extensions of the
$SG$ with a $Z_2$ IGG. Here $SG$ is the direct product of the space
group of honeycomb lattice and the time-reversal $Z_2$ group. In
appendix \ref{app:PSG General conditions} we show the general
constraints that must be satisfied for such a group extension. In
appendix \ref{app:160Z2}, using these constraints, we find there are
in total 160 $Z_2$ algebraic-PSGs on honeycomb lattice. And at most 128
PSGs of them can be realized by an ansatz $\{u_{ij}\}$. These 128
PSGs are the complete classification of $Z_2$ spin liquids on a
honeycomb lattice.

\section{Classification of $Z_2$ states around the u-RVB
state}\label{Z2_PSG}

Can one further identify the candidate states for the spin liquid
discovered in the numerical study\cite{Meng2010}? The answer is yes.
Numerically the spin liquid phase is found close to the Mott
transition and it seems to be connected to the semimetal phase by a
continuous phase transition. What are the $Z_2$ Schwinger-fermion
states in the neighborhood of the semi-metal phase?

Are there Schwinger-fermion mean-field states that can be connected
to the semi-metal phase via a continuous phase transition? This
question was firstly discussed by Hermele in \Ref{hermele:035125}.
Using slave-rotor formalism, it was shown that the semi-metal phase
can go through a continuous phase transition into an $SU(2)$ u-RVB
state (also termed as algebraic spin liquid (ASL) in
\Ref{hermele:035125}) at the mean-field level. This $SU(2)$ u-RVB
ansatz, in terms of $f$-spinon, can be written as
$H_{MF}=t\sum_{<ij>}f_{i\alpha}^{\dagger}f_{j\alpha}$, $t$ is real
and summation is over all nearest neighbor bond. The single-spinon
dispersion of u-RVB state is similar to the electronic dispersion in
the semi-metal phase, which is composed of four two-component Dirac
cones at the corner of Brillouin Zone, two from spin and two from
valley. Physically it is easy to understand u-RVB state connecting
with the semi-metal phase: At the Mott transition, only the charge
fluctuation becomes fully gapped and the spinon dispersion still
remember the semi-metal band structure.

The u-RVB ansatz can be simply expressed as a graphene-like nearest neighbor hopping of $f$-fermions:
Fig.\ref{fig:honeycomb}:
\begin{align}
 H_{MF}^{uRVB}&=\chi\sum_{<ij>}f_{i\alpha}^{\dagger}f_{j\alpha},\label{eq:urvb}
\end{align}
where $\chi$ is real. Beyond mean-field level, the low-energy
effective theory of u-RVB state is described by $N_f=2$~
two-component Dirac spinons ($SU(2)$ gauge doublet) coupled with a
dynamical $SU(2)$ gauge field\cite{hermele:035125}, \ie QCD$_3$. In
the large-$N_f$ limit QCD$_3$ has a stable IR fixed point with
gapless excitations and can be a stable ASL phase\cite{3622404}.
When $N_f=0$ the pure gauge QCD$_3$ is in a confined
phase\cite{PhysRevLett.30.1343,PhysRevLett.30.1346}. This indicates
a critical $N_c$ and when $N_f<N_c$ confinement
occurs\cite{3622404}. Although no controlled estimate of $N_c$ is
available, a self-consistent solution of the Schwinger-Dyson
equations\cite{3622404} suggests $N_c\approx \frac{64}{\pi^2}$. We
will assume that $N_c>2$ and therefore u-RVB state is not a stable
phase.

\begin{figure}
 \includegraphics[width=0.3\textwidth]{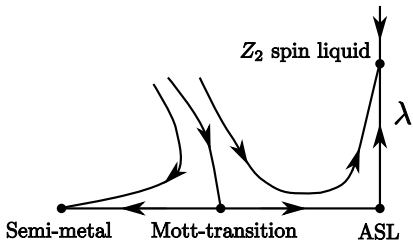}
\caption{Schematic RG flow of the Mott transition. $\lambda$
represents a relevant perturbation of the ASL fixed point, which
eventually drive the RG flow into a stable $Z_2$ spin liquid fixed
point.} \label{fig:rg}
\end{figure}

Due to the lack of the knowledge of the confinement mechanism, it is
difficult to reliably predict the ultimate fate of the u-RVB state
(or ASL). But one possibility is that the strong gauge interaction
induces Higgs condensation which breaks the $SU(2)$ gauge symmetry
down to $Z_2$, so that the renormalization group flows into a stable
fixed point of $Z_2$ gauge theory. Based on this assumption, we can
propose a scenario of a continuous phase transition from the
semi-metal phase into a $Z_2$ spin liquid phase: the critical point
is still described by the slave-rotor critical theory discussed in
\Ref{hermele:035125}. But on the Mott insulator side, a dangerously
irrelevant operator (for example, can be a four-fermion interaction
term) becomes relevant and finally drive the RG flow away from the
ASL fixed point and flow into a stable fixed point of a $Z_2$ phase
by Higgs mechanism. Here we assume the ASL still describes an
unstable fixed point with relevant directions. This scenario is
schematically shown in Fig.\ref{fig:rg}, which has the same spirit
of the deconfined quantum criticality\cite{senthil-2004-303}.

If this scenario is correct, the mean-field ansatz of the $Z_2$ spin
liquid should be connected to the u-RVB ansatz by a continuous Higgs
condensation, which breaks the $SU(2)$ IGG down to $Z_2$. During
this transition, the u-RVB ansatz $\{u_{ij}^{uRVB}\}\rightarrow
\{u_{ij}^{uRVB}+\delta u_{ij}\}$ and the $\delta u_{ij}$ amplitudes
play the role of the Higgs boson. We define a $Z_2$ state to be
around (or in the neighborhood of) the u-RVB when the $Z_2$ state
can be obtained by an infinitesimal change
$\{u_{ij}^{uRVB}\}\rightarrow \{u_{ij}^{uRVB}+\delta u_{ij}\}$.

The PSG of $\{u_{ij}^{uRVB}+\delta u_{ij}\}$ must be a subgroup of
the PSG of the u-RVB state Eq.(\ref{eq:urvb}). In appendix
\ref{app:24Z2_uRVB} we classify all these possible PSG subgroups
with the $Z_2$ IGG, which allows us to construct all possible $Z_2$
states around the u-RVB state. This technique was firstly developed
by Wen\cite{PhysRevB.65.165113}. We find 24 gauge inequivalent $Z_2$ PSGs as listed
in Table \ref{tab:z2_urvb} in appendix \ref{app:24Z2_uRVB}.

Can these 24 $Z_2$ SL states have a full energy gap? We find not all
of them can have a gapped spinon spectrum. This can be understood
starting from a Dirac dispersion of the u-RVB state. To gap out the
Dirac nodes, at least one mass term in the low-energy effective
theory of a given $Z_2$ state must be allowed by symmetry. In
appendix \ref{app:mass} we show that only 4 of the 24 $Z_2$ states
allow mass term in the low energy theory. Thus only these 4 states
are fully gapped $Z_2$ spin liquids around u-RVB state. The other 20
states have symmetry protected gapless spinon dispersions.

These four states are state \#16,\#17,\#19, and \#22 in Table
\ref{tab:z2_urvb} in appendix \ref{app:24Z2_uRVB}. We can generate
their mean-field ansatzs by these PSGs. We find that up to the 3rd
neighbor mean-field amplitudes $u_{(\alpha,\beta,\gamma)}$ as shown
in Fig.\ref{fig:honeycomb}, only one of these four states can be
realized, which is state \#19. As shown in appendix
\ref{app:mass:4_z2}, mean-field ansatzs up to the 3rd neighbor of
the other three states actually have a $U(1)$ IGG. Only after
introducing longer-range mean-field bonds can these three states
have a $Z_2$ IGG. In particular, state \#16 requires 5th neighbor,
state \#17 requires 4th neighbor and state \#22 requires 9th
neighbor amplitudes, while state \#19 only requires 2nd neighbor
amplitudes. Because the $t/U$ expansion of the Hubbard model give a
rather short-ranged spin interaction for the SL phase found in
numerics\cite{Meng2010} ($t/U\sim 1/4$), the other three states are
unlikely to be realized in a Hubbard model on honeycomb lattice.

After choosing a proper gauge, the mean-field ansatz of \#19 can be
expressed as a sublattice dependent pairing of the $f$-spinons, as
shown in Fig.\ref{fig:honeycomb}:
\begin{align}
H_{MF}=&\chi\sum_{<ij>}f_{i\alpha}^{\dagger}f_{j\alpha}+\Delta e^{i\theta}\sum_{<<ij>>\in A} \epsilon_{\alpha\beta}f_{i\alpha}^{\dagger}f_{j\beta}^\dagger\notag\\
&+\Delta e^{-i\theta}\sum_{<<ij>>\in B}
\epsilon_{\alpha\beta}f_{i\alpha}^{\dagger}f_{j\beta}^\dagger+\text{h.c.}
\label{eq:SPS}
\end{align}
and we term it as sublattice pairing state (SPS). Note that
$\theta\neq 0,\pm\pi/2,\pi$, because otherwise the ansatz has $U(1)$
IGG. We propose SPS to be the SL phase found in numerics.

\section{Schwinger-fermion mean-field study of the $J_1-J_2$ model on honeycomb lattice}\label{J1J2MF}
\begin{figure}
\includegraphics[width=0.45\textwidth]{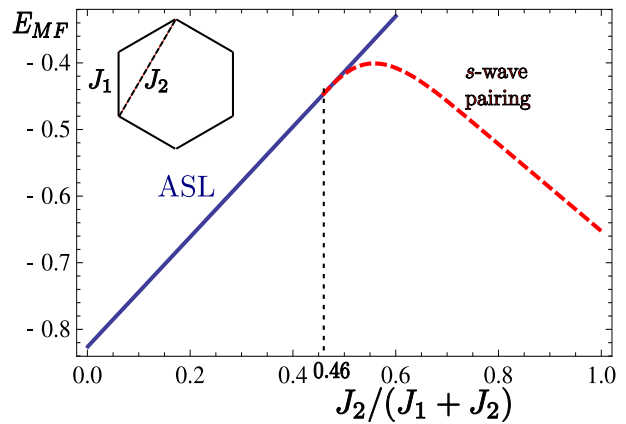}
\caption{Mean-field phase diagram of $J_1-J_2$ model by Schwinger-fermion approach.}
\label{fig:J1J2MF}
\end{figure}

Can SPS be realized in the Hubbard model when $t/U\sim 1/4$, where
numerics shows a gapped SL phase? In particular, by the Mott
transition theory of Hermele\cite{hermele:035125}, the u-RVB (or
ASL) state is in the neighborhood of the Mott transition. Can SPS be
more favorable than the ASL state? To address this question, we use
$t/U$ expansion of the Hubbard model\cite{PhysRevB.37.9753} to
obtain an effective $J_1-J_2$ spin model on honeycomb lattice:
\begin{align}
H=J_1\sum_{<ij>}\vec S_i\cdot S_j+J_2\sum_{<<ij>>}\vec S_i\cdot S_j
\end{align}
 where $J_1$ and $J_2$ are the 1st neighbor and 2nd neighbor antiferromagnetic coupling.
 Following \Ref{PhysRevB.37.9753}, we find up to $t^4/U^3$ order, the effective $J_1$ and $J_2$ are:
\begin{align}
 J_1&=4t^2/U-16t^4/U^3,& J_2&=4t^4/U^3.
\end{align}
Naively plugging in $t/U\sim 1/4$ gives $J_2/J_1\sim 1/12$.

We use the variationally mean-field ansatz Eq.\ref{eq:SPS}. Note that this mean-field study is biased towards spin disordered ground state. For example, we do not include Neel order which is known to be the ground state at $J_2=0$, and we also do not include the spiral spin order which is found by semiclassical study of $J_1-J_2$ model\cite{Rastelli19791,Fouet_J1J2J3}.  The purpose of the current mean-field study is to understand whether a gapped spin liquid can be more favorable compared to the gapless ASL state when $J_2$ is tuned up and frustration becomes important.

By minimizing the mean-field energy in Eq.(\ref{eq:action}), the phase
diagram of $J_1-J_2$ model is obtained and shown in Fig.\ref{fig:J1J2MF}, where
we fix $J_1+J_2=1$ and $E_{MF}$ is scaled from Eq.(\ref{eq:action})
by $8/3$. We find that when $J_2/J_1<0.85$ (or $J_2/(J_1+J_2)<0.46$), the ground state is
u-RVB(or ASL) state: $\chi\neq0$ and $\Delta=0$. When
$J_2/J_1>0.85$, the ground state is an $s$-wave pairing state:
$\chi,\Delta\neq0$ and $\theta=0$. The $s$-wave pairing state opens
an energy gap for spinons but has remaining $U(1)$ gapless gauge
fluctuation. Due to monopole proliferation\cite{Polyakov} the $s$-wave
pairing state is not a stable phase. In this mean-field study, the
gauge fluctuations are not considered and this is the reason why we
find $s$-wave pairing state as a ground state. Taking gauge
fluctuations into account, the likely fate of the $s$-wave pairing
state is that $\theta$ becomes nonzero and the $Z_2$ SPS state is
realized.

We propose to study the $J_1-J_2$ model by Gutzwiller projected
wavefunction variational approach\cite{3381199} because it can be
viewed as a method to include the gauge fluctuation. We leave this
projected wavefunction study as a direction of future research,
which may realize SPS as the ground state. Projected wavefunctions
are also classified by PSG, so the present work also provide
guideline for the search of ground states in the projected
wavefunction space.

\section{Discussion}
In this work we completely classified the $Z_2$ mean-field states in
the Schwinger-fermion approach. Using physical argument, we identify
a single state: SPS, as the possible spin liquid phase found in the
recent Quantum Monte Carlo study of the Hubbard model on a honeycomb
lattice\cite{Meng2010}. SPS is in the neighborhood of the semimetal
phase and we propose a scenario for the continuous transition
connecting the two phases.

In our mean-field study of the $J_1-J_2$ model, the $s$-wave pairing
state is realized for a fairly large $J_2$, corresponding to a
fairly large $t/U\sim 0.44$. A higher order spin-spin effective
interaction such as the 6-spin ring exchange term and/or a more
careful projected wavefunction study may realize SPS phase for a
smaller $t/U$.

In a recent work\cite{Fa_honeycomb_PSG}, Wang study the $Z_2$ mean-field states in
the Schwinger-boson approach, and identify a zero-flux SL state,
which is naturally connected to a Neel ordered state by a
potentially continuous phase transition. Whether the SPS found in
the present work is related to Wang's result is unclear. And we
leave the possible continuous transition from SPS to the Neel
ordered phase as a subject of future research.

YR thanks Ashvin Vishwanath and Fa Wang for helpful discussions. YML
thanks Prof. Ziqiang Wang for support during this work under DOE
Grant DE-FG02-99ER45747. YR is supported by the start-up fund at
Boston College.

\appendix

\section{General conditions on projective symmetry groups on a honeycomb
lattice}\label{app:PSG General conditions}

As mentioned in section \ref{SF_PSG}, SG on a honeycomb lattice is
generated by time reversal transformation $\boldsymbol T$,
translations along $\vec a_1,\vec a_2$: $T_1,T_2$,
plaquette-centered 60$^\circ$ $C_6$ rotation, and a horizontal
mirror reflection $\boldsymbol{\sigma}$ as shown in
Fig.\ref{fig:honeycomb}. In the present problem, the symmetry group
$SG$ can be represented as
\begin{eqnarray}\nonumber
SG=\{U=\bs{T}^{\nu_{\bs{T}}}\cdot T_1^{\nu_{T_1}}\cdot
T_2^{\nu_{T_2}}\cdot
C_6^{\nu_{C_6}}\cdot\bs{\sigma}^{\nu_{\bs{\sigma}}}\}
\end{eqnarray}
where $\nu_{T_1},\nu_{T_2}\in\bs{Z}$ and
$\nu_{\bs{T}},\nu_\bs{\sigma}\in\bs{Z_2}$,~$\nu_{C_6}\in\bs{Z_6}$,
since the generators satisfy
\begin{eqnarray}
\bs{T}^2=\bs{\sigma}^2=(C_6)^6=1\label{T^2|S^2|C_6^2}
\end{eqnarray}
Here $1$ stands for the identity element of $SG$. To completely
determine the multiplication rule of this group, we need to identify
the multiplication rule of two different generators in an order
different from $\bs{T}^{\nu_{\bs{T}}}\cdot T_1^{\nu_{T_1}}\cdot
T_2^{\nu_{T_2}}\cdot
C_6^{\nu_{C_6}}\cdot\bs{\sigma}^{\nu_{\bs{\sigma}}}$:
\begin{eqnarray}
&U\bs{T}=\bs{T}U~~~(U=T_1,T_2,C_6,\bss)\\
&T_1T_2=T_2T_1\\
&C_6T_1=T_2C_6\\
&C_6T_2=T_1^{-1}T_2C_6\\
&\bs{\sigma}T_1=T_1\bs{\sigma}\\
&\bss T_2=T_1T_2^{-1}\bss\\
&\bss C_6=C_6^{-1}\bss\label{sig C6}
\end{eqnarray}

The above relations can be written in an alternative way
\begin{eqnarray}
&\bs{T}^2=\bs{\sigma}^2=(C_6)^6=1\label{T^2|S^2|C6^2}\\
&\bs{T}U\bs{T}^{-1}U^{-1}=1~~~(U=T_1,T_2,C_6,\bss)\label{T-U}\\
&T_1T_2T_1^{-1}T_2^{-1}=1\label{T2-T1}\\
&T_2^{-1}C_6T_1C_6^{-1}=1\label{C6-T1}\\
&T_1^{-1}C_6T_1T_2^{-1}C_6^{-1}=1\label{C6-T2}\\
&T_1^{-1}\bss T_1\bss^{-1}=1\label{sig-T1}\\
&T_2^{-1}\bss T_1T_2^{-1}\bss^{-1}=1\label{sig-T2}\\
&\bss C_6\bss C_6=1\label{sig-C6}
\end{eqnarray}
which determines the inverse of all the group elements.

As introduced in section \ref{SF_PSG}, the mean-field ansatz
$\{u_{ij}\}$ of a spin liquid is invariant under the action of any
element $G_UU$ of a projective symmetry group (PSG). The
multiplication rule of the symmetry group would immediately enforce
the following constraints on a PSG by its definition: if
$U_1U_2=U_3$ then
\begin{eqnarray}
&G_{U_1}U_1G_{U_2}U_2(\{u_{ij}\})=G_{U_3}U_3(\{u_{ij}\})\Longrightarrow\notag\\
&\big[G_{U_1}\big(U_1U_2(i)\big)G_{U_2}\big(U_2(i)\big)\big]u_{ij}\big[G_{U_1}\big(U_1U_2(i)\big)G_{U_2}\big(U_2(i)\big)\big]^\dagger\notag\\
&=\big[G_{U3}\big(U_3(i)\big)]u_{ij}\big[G_{U3}\big(U_3(i)\big)\big]^\dagger,~~~\forall
~i,j\label{generic constraint_psg}
\end{eqnarray}
On the other hand, we know those pure gauge transformations, under
which the mean-field ansatz $\{u_{ij}\}$ is invariant, constitute a
subgroup of PSG, coined the invariant gauge group (IGG):
\begin{eqnarray}\label{igg}
IGG=\{W_i|W_iu_{ij}W_j^\dagger=u_{ij},~~W_i\in SU(2)\}
\end{eqnarray}
Therefore from (\ref{generic constraint_psg}) we have the following
constraints on the elements of a PSG
\begin{eqnarray}
\big[G_{U_1U_2}\big(U_1U_2(i)\big)\big]^\dagger
G_{U_1}\big(U_1U_2(i)\big)G_{U_2}\big(U_2(i)\big)=\mathcal{G}\in
IGG\notag
\end{eqnarray}
The above condition holds for any two group elements $U_1,U_2$ of
SG. Similar with SG, we can choose a set of generators in any given
PSG:
$\{G_{T_1}T_1,G_{T_2}T_2,G_{\boldsymbol{T}}\boldsymbol{T},G_{C_6}C_6,G_{\boldsymbol{\sigma}}\boldsymbol{\sigma}\}$.
Any given element in PSG can be written in the standard form:
\begin{align}
 G_UU=&(G_{\boldsymbol{T}}\boldsymbol T)^{\nu_{\boldsymbol T}} \cdot (G_{T_1}T_1)^{\nu_{T_1}}\cdot (G_{T_2}T_2)^{\nu_{T_2}}\notag\\
& \cdot (G_{C_6}C_6)^{\nu_{C_6}}\cdot
(G_{\boldsymbol{\sigma}}\boldsymbol{\sigma})^{\nu_{\boldsymbol{\sigma}}}
\end{align}
Since the multiplication rule of SG on a honeycomb lattice is
completely determined by (\ref{T^2|S^2|C_6^2})-(\ref{sig C6}), or
equivalently (\ref{T^2|S^2|C6^2})-(\ref{sig-C6}), the only
independent constraints on the PSG generators are the following:
\begin{eqnarray}
&(G_{\boldsymbol{T}}\boldsymbol{T})^2\in IGG\\
&(G_{\boldsymbol\sigma}\boldsymbol{\sigma})^2\in IGG\notag\\
&(G_{C_6}C_6)^6\in IGG\notag\\
&(G_{T_1}T_1)^{-1}(G_{T_2}T_2)^{-1}(G_{T_1}T_1)(G_{T_2}T_2)\in IGG\notag\\
&(G_{T_1}T_1)^{-1}(G_{\boldsymbol{T}}\boldsymbol{T})^{-1}(G_{T_1}T_1)(G_{\boldsymbol T}\boldsymbol{T})\in IGG\notag\\
&(G_{T_2}T_2)^{-1}(G_{\boldsymbol T}\boldsymbol{T})^{-1}(G_{T_2}T_2)(G_{\boldsymbol T}\boldsymbol{T})\in IGG\notag\\
&(G_{T_2}T_2)^{-1}(G_{C_6}C_6)(G_{T_1}T_1)(G_{C_6}C_6)^{-1}\in IGG\notag\\
&(G_{T_1}T_1)^{-1}(G_{C_6}C_6)(G_{T_1}T_1)(G_{T_2}T_2)^{-1}(G_{C_6}C_6)^{-1}\in IGG\notag\\
&(G_{\boldsymbol T}\boldsymbol{T})^{-1}(G_{C_6}C_6)^{-1}(G_{\boldsymbol T}\boldsymbol{T})(G_{C_6}C_6)\in IGG\notag\\
&(G_{T_1}T_1)^{-1}(G_{\boldsymbol\sigma}\boldsymbol{\sigma})(G_{T_1}T_1)(G_{\boldsymbol{\sigma}}\boldsymbol{\sigma})^{-1}\in IGG\notag\\
&(G_{T_2}T_2)^{-1}(G_{\boldsymbol\sigma}\boldsymbol{\sigma})(G_{T_1}T_1)(G_{T_2}T_2)^{-1}(G_{\boldsymbol\sigma}\boldsymbol{\sigma})^{-1}\in IGG\notag\\
&(G_{\boldsymbol\sigma}\boldsymbol{\sigma})(G_{C_6}C_6)(G_{\boldsymbol{\sigma}}\boldsymbol{\sigma})(G_{C_6}C_6)\in IGG\notag\\
&(G_{\boldsymbol
T}\boldsymbol{T})^{-1}(G_{\boldsymbol\sigma}\boldsymbol{\sigma})^{-1}(G_{\boldsymbol
T}\boldsymbol{T})(G_{\boldsymbol\sigma}\boldsymbol{\sigma})\in
IGG\notag
\end{eqnarray}
or more specifically
\begin{eqnarray}
&\big[G_{\boldsymbol{T}}(i)\big]^2\in IGG,\\
&G_{\boldsymbol\sigma}\big(\bss(i)\big)G_{\bss}(i)\in IGG,\notag\\
&G_{C_6}\big(C_6^{-1}(i)\big)G_{C_6}\big(C_6^{-2}(i)\big)G_{C_6}\big(C_6^{3}(i)\big)\notag\\
&\cdot
G_{C_6}\big(C_6^{2}(i)\big)G_{C_6}\big(C_6(i)\big)G_{C_6}(i)\in IGG,\notag\\
&G_{T_1}^{-1}\big(T_2^{-1}T_1(i)\big)G_{T_2}^{-1}\big(T_1(i)\big)G_{T_1}\big(T_1(i)\big)G_{T_2}(i)\in IGG,\notag\\
&G_{T_1}^{-1}\big(T_1(i)\big)G_{\boldsymbol{T}}^{-1}\big(T_1(i)\big)G_{T_1}\big(T_1(i)\big)G_{\boldsymbol T}(i)\in IGG,\notag\\
&G_{T_2}^{-1}\big(T_2(i)\big)G_{\boldsymbol T}^{-1}\big(T_2(i)\big)G_{T_2}\big(T_2(i)\big)G_{\boldsymbol T}(i)\in IGG,\notag\\
&G_{T_2}^{-1}\big(T_2(i)\big)G_{C_6}\big(T_2(i)\big)G_{T_1}\big(T_1C_6^{-1}(i)\big)G_{C_6}^{-1}(i)\in IGG,\notag\\
&G_{T_1}^{-1}\big(T_1(i)\big)G_{C_6}\big(T_1(i)\big)G_{T_1}\big(C_6^{-1}T_1(i)\big)\notag\\
&\cdot G_{T_2}^{-1}\big(C_6^{-1}(i)\big)G_{C_6}^{-1}(i)\in IGG,\notag\\
&G_{\boldsymbol T}^{-1}\big(C_6^{-1}(i)\big)G_{C_6}^{-1}(i)G_{\boldsymbol T}(i)G_{C_6}(i)\in IGG,\notag\\
&G_{T_1}^{-1}\big(T_1(i)\big)G_{\boldsymbol\sigma}\big(T_1(i)\big)G_{T_1}\big(T_1\bss^{-1}(i)\big)G_{\boldsymbol{\sigma}}^{-1}(i)\in IGG,\notag\\
&G_{T_2}^{-1}\big(T_2(i)\big)G_{\boldsymbol\sigma}\big(T_2(i)\big)G_{T_1}\big(\bss T_2(i)\big)G_{T_2}^{-1}\big(\bss(i)\big)G_{\boldsymbol\sigma}^{-1}(i)\in IGG,\notag\\
&G_{\boldsymbol\sigma}(i)G_{C_6}\big(\bss(i)\big)G_{\boldsymbol{\sigma}}\big(\bss C_6(i)\big)G_{C_6}\big(C_6(i)\big)\in IGG,\notag\\
&G_{\boldsymbol
T}^{-1}(\bss(i))G_{\boldsymbol\sigma}^{-1}(i)G_{\boldsymbol
T}(i)G_{\boldsymbol\sigma}(i)\in IGG.\notag
\end{eqnarray}
Above are all the general consistent conditions to be satisfied by
the generators of a PSG on a honeycomb lattice.

We will use $(x_1,x_2,s)$ to label a site $i$ in a honeycomb lattice, where
$x_1,x_2$ are the coordinates of the unit cell in basis $\vec
a_1,\vec a_2$ and $s=0,1$ for $A$ and $B$ sublattice respectively.
 For convenience, we summarize the coordinate transformation of all
the generators in the symmetry group on a honeycomb lattice as
follows:
\begin{eqnarray}
&\bs{T}:~~~(x_1,x_2,s)\rightarrow(x_1,x_2,s),\label{sg_transf}\\
&T_1:~~~(x_1,x_2,s)\rightarrow(x_1+1,x_2,s),\notag\\
&T_2:~~~(x_1,x_2,s)\rightarrow(x_1,x_2+1,s),\notag\\
&\bss:~~~(x_1,x_2,s)\rightarrow(x_1+x_2,-x_2,1-s),\notag\\
&C_6:~~~(x_1,x_2,0)\rightarrow(1-x_2,x_1+y_1-1,1)\notag\\
&~~~~~~~(x_1,x_2,1)\rightarrow(-x_2,x_1+y_1,0)\notag
\end{eqnarray}\\

\section{Classification of $Z_2$ projective symmetry
groups on a honeycomb lattice}\label{app:160Z2}

As discussed in section \ref{SF_PSG}, the problem of classifying all
possible $Z_2$ Schwinger-fermion mean-field states is mathematically
reduced to finding all possible PSGs. Let us firstly find all
algebraic PSGs.

\subsection{General discussions}

In the case of $Z_2$ spin liquids, the IGG of the corresponding PSG
is a $Z_2$ group: $IGG=\{\pm\tau^0\}$. The constraints listed in
appendix \ref{app:PSG General conditions} now becomes
\begin{eqnarray}
&\big[G_{\boldsymbol{T}}(i)\big]^2=\eta_{\bs T}\tau^0,\label{T^2}\\
&G_{\boldsymbol\sigma}\big(\bss(i)\big)G_{\bss}(i)=\eta_\bss \tau^0,\label{sig^2}\\
&G_{C_6}\big(C_6^{-1}(i)\big)G_{C_6}\big(C_6^{-2}(i)\big)G_{C_6}\big(C_6^{3}(i)\big)\\
&\cdot
G_{C_6}\big(C_6^{2}(i)\big)G_{C_6}\big(C_6(i)\big)G_{C_6}(i)=\eta_{C_6}\tau^0,\label{c6^6}\\
&G_{T_1}^{-1}\big(T_2^{-1}T_1(i)\big)G_{T_2}^{-1}\big(T_1(i)\big)\notag\\
&\cdot G_{T_1}\big(T_1(i)\big)G_{T_2}(i)=\eta_{12}\tau^0,\label{T1-T2}\\
&G_{T_1}^{-1}\big(T_1(i)\big)G_{\boldsymbol{T}}^{-1}\big(T_1(i)\big)G_{T_1}\big(T_1(i)\big)G_{\boldsymbol T}(i)=\eta_{1\bs T}\tau^0,\label{T1_T}\\
&G_{T_2}^{-1}\big(T_2(i)\big)G_{\boldsymbol T}^{-1}\big(T_2(i)\big)G_{T_2}\big(T_2(i)\big)G_{\boldsymbol T}(i)=\eta_{2\bs T}\tau^0,\label{T2_T}\\
&G_{T_2}^{-1}\big(T_2(i)\big)G_{C_6}\big(T_2(i)\big)\notag\\
&\cdot G_{T_1}\big(T_1C_6^{-1}(i)\big)G_{C_6}^{-1}(i)=\eta_{C_61}\tau^0,\label{T2-c6}\\
&G_{T_1}^{-1}\big(T_1(i)\big)G_{C_6}\big(T_1(i)\big)G_{T_1}\big(C_6^{-1}T_1(i)\big)\notag\\
&\cdot G_{T_2}^{-1}\big(C_6^{-1}(i)\big)G_{C_6}^{-1}(i)=\eta_{C_62}\tau^0,\label{T1-c6}\\
&G_{\boldsymbol T}^{-1}\big(C_6^{-1}(i)\big)G_{C_6}^{-1}(i)G_{\boldsymbol T}(i)G_{C_6}(i)=\eta_{C_6\bs T}\tau^0,\label{T-c6}\\
&G_{T_1}^{-1}\big(T_1(i)\big)G_{\boldsymbol\sigma}\big(T_1(i)\big)\notag\\
&\cdot G_{T_1}\big(T_1\bss^{-1}(i)\big)G_{\boldsymbol{\sigma}}^{-1}(i)=\eta_{\bss1}\tau^0,\label{T1-sig}\\
&G_{T_2}^{-1}\big(T_2(i)\big)G_{\boldsymbol\sigma}\big(T_2(i)\big)G_{T_1}\big(\bss T_2(i)\big)\notag\\
&\cdot G_{T_2}^{-1}\big(\bss(i)\big)G_{\boldsymbol\sigma}^{-1}(i)=\eta_{\bss2}\tau^0,\label{T2-sig}\\
&G_{\boldsymbol\sigma}(i)G_{C_6}\big(\bss(i)\big)\notag\\
&\cdot G_{\boldsymbol{\sigma}}\big(\bss C_6(i)\big)G_{C_6}\big(C_6(i)\big)=\eta_{\bss C_6}\tau^0,\label{c6-sig}\\
&G_{\boldsymbol
T}^{-1}(\bss(i))G_{\boldsymbol\sigma}^{-1}(i)G_{\boldsymbol
T}(i)G_{\boldsymbol\sigma}(i)=\eta_{\bss \bs T}\tau^0.\label{T-sig}
\end{eqnarray}
where all the $\eta$'s take value of $\pm1$. Not all of these
conditions are gauge independent. Because we can re-choose the gauge
part of generators such as $G_{T_1},G_{T_2}...$ by multiplying them
by $-\tau^0$ (an element of IGG), only those conditions in which the
same generator shows up twice are gauge independent. We can use this
gauge dependence to simplify these conditions. Because
$G_{T_1}$($G_{T_2}$) only show up once in the equation of
$\eta_{C_61}$($\eta_{C_6 2}$), we can always choose a gauge such
that $\eta_{C_61}=\eta_{C_62}=1$. All other $\eta$'s are gauge
independent.

In the following we will determine all the possible PSG's with
different (gauge inequivalent) elements $\{G_U(i)\}$. These
different PSG's characterize all the different type of $Z_2$ spin
liquids on a honeycomb lattice, which might be constructed from
mean-field ansatz $\{u_{ij}\}$.

First notice that under a local $SU(2)$ gauge transformation
$u_{ij}\rightarrow W_iu_{ij}W_j^\dagger$, the PSG elements transform
as $G_U(i)\rightarrow W_iG_U(i)W^\dagger_{U^{-1}(i)}$. Making use of
such a degree of freedom, we can always choose proper gauge so that
\begin{eqnarray}
G_{T_1}(x_1,x_2,s)=G_{T_2}(0,x_2,s)=\tau^0,~~~x_1,x_2\in\mathbb{Z}.\notag
\end{eqnarray}
Now taking (\ref{T1-T2}) into account, we have
$G_{T_2}(\{x_1+1,x_2,s\})=\eta_{12}G_{T_2}(\{x_1,x_2,s\})$ and
therefore
\begin{eqnarray}
&G_{T_1}(x_1,x_2,s)=\tau^0\\
&G_{T_2}(x_1,x_2,s)=\eta_{12}^{x_1}\tau^0\notag
\end{eqnarray}

Meanwhile, from (\ref{T^2}), (\ref{T1_T}) and (\ref{T2_T}) we can
immediately see that $\eta_{1\bs T}=\eta_{2\bs T}=1$, and the gauge
inequivalent choices of $G_{\bs T}(i)$ are the following
\begin{eqnarray}
G_{\bs T}(x_1,x_2,s)=g_{\bs
T}(s)=\left.\Big\{\begin{aligned}\eta_t^s\tau^0,~~~&\eta_{\bs
T}=1\\
i\tau^3,~~~&\eta_{\bs T}=-1\end{aligned}\right.
\end{eqnarray}
where $\eta_t=\pm1$.

As discussed earlier, we can always choose a proper gauge so that
$\eta_{C_61}=\eta_{C_62}=1$. Then from conditions (\ref{T2-c6}) and
(\ref{T1-c6}) we see that
\begin{eqnarray}\label{g_c6}
G_{C_6}(x_1,x_2,s)=\eta_{12}^{x_1x_2+x_1(x_1-1)/2}g_{C_6}(s)
\end{eqnarray}
similarly from conditions (\ref{T1-sig}) and (\ref{T2-sig}) we have
\begin{eqnarray}\label{g_sig}
G_\bss(x_1,x_2,s)=\eta_{\bss1}^{x_1}\eta_{\bss2}^{x_2}\eta_{12}^{x_2(x_2-1)/2}g_\bss(s)
\end{eqnarray}
where $g_{C_6}(s),g_\bss(s)\in SU(2)$. Note that (\ref{sig^2}) and
(\ref{c6-sig}) give further constraints to the above expression
(\ref{g_sig}):
\begin{eqnarray}
\eta_{\bss1}=\eta_{\bss2}=\eta_{12}
\end{eqnarray}
Now we see the elements of PSG can be expressed as
\begin{eqnarray}
&G_{T_1}(x_1,x_2,s)=\tau^0\\
&G_{T_2}(x_1,x_2,s)=\eta_{12}^{x_1}\tau^0\notag\\
&G_{\bs T}(x_1,x_2,s)=g_{\bs T}(s)\\
&G_{C_6}(x_1,x_2,s)=\eta_{12}^{x_1x_2+x_1(x_1-1)/2}g_{C_6}(s)\notag\\
&G_\bss(x_1,x_2,s)=\eta_{12}^{x_1+x_2(x_2+1)/2}g_\bss(s)\notag
\end{eqnarray}
Consistent conditions (\ref{sig^2}), (\ref{c6^6}), (\ref{T-c6}),
(\ref{c6-sig}) and (\ref{T-sig}) correspond to the following
constraints on $SU(2)$ matrices $g_{C_6}(s),~g_\bss(s)$:
\begin{eqnarray}\label{g_constraint}
&g_\bss(0)g_\bss(1)=\eta_{\bss}\tau^0,\\
&[g_{C_6}(s)g_{C_6}(1-s)]^3=\eta_{C_6}\eta_{12}\tau^0,\notag\\
&g_{\bs T}(s)g_{C_6}(s)=g_{C_6}(s)g_{\bs T}(1-s)\eta_{C_6\bs T}\notag\\
&g_{\bs T}(s)g_\bss(s)=g_\bss(s)g_{\bs T}(1-s)\eta_{\bss\bs T}\notag\\
&g_\bss(s)g_{C_6}(1-s)=\left.\Big\{\begin{aligned}\lambda_{C_6}^s\tau^0,~~~&\eta_{\bss
C_6}=1\\
\imth\hat{n}_s\cdot\vec\tau,~~~&\eta_{\bss
C_6}=-1\end{aligned}\right.\notag
\end{eqnarray}
where $\lambda_{C_6}=\pm1$ and $\hat{n}_s$ is a unit vector.

\subsection{A summary of 160 different PSG's}

Below we summarize all the 160 possible PSG's obtained through
solving (\ref{g_constraint}). We use capital Roman numerals
(\Rmnum1) and (\Rmnum2) to label $g_{\bs T}=\eta_t^s\tau^0$ and
$g_{\bs T}=i\tau^3$ respectively. Roman numerals (\rmnum1) and
(\rmnum2) are used to label $\eta_{C_6\bs T}=\pm1$ respectively. (A)
and (B) are used to label $\eta_{\bss C_6}=\pm1$ respectively.
Finally ($\alpha$) and ($\beta$) are used to label $\eta_{\bss\bs
T}$ respectively.\\

(\Rmnum{1})~~~$g_{\bss T}=\eta_t^s\tau^0$:

It's easy to see that $\eta_{C_6\bs T}=\eta_{\bss\bs T}=\eta_t$ from
(\ref{g_constraint}), so there is the only possibility among
(\rmnum1) and (\rmnum2).

(A)~~~$g_\bss(s)=\lambda_{C_6}^sg_{C_6}^{-1}(1-s)$:

we have $\lambda_{C_6}=\eta_\bss\eta_{C_6}\eta_{12}$ and
\begin{eqnarray}
&g_{C_6}(0)=\tau^0,\\
&g_{C_6}(1)=g_\bss(0)=\eta_{C_6}\eta_{12}\tau^0\notag,\\
&g_\bss(1)=\eta_\bss\eta_{C_6}\eta_{12}\tau^0.\notag
\end{eqnarray}
This represents $2^4=\alert{16}$ different PSG's in the class
(\Rmnum1)(A) since $\eta_t,\eta_{C_6},\eta_\bss,\eta_{12}=\pm1$.

(B)~~~$g_\bss(s)g_{C_6}(1-s)=\imth\hat{n}_s\cdot\vec\tau$:

Choosing a proper gauge (so that $g_{C_6}(0)=\tau^0$) we have
\begin{eqnarray}
&g_{C_6}(0)=\tau^0,\\
&g_{C_6}(1)=\eta_{C_6}\eta_{12}e^{\imth\psi_3\tau^3}\notag,\\
&g_\bss(0)=\imth\tau^1\eta_{C_6}\eta_{12}e^{-\imth\psi_3\tau^3}\notag,\\
&g_\bss(1)=-\imth\eta_\bss\eta_{C_6}\eta_{12}e^{\imth\psi_3\tau^3}\tau^1\notag.
\end{eqnarray}
where $\psi_3\equiv0,\pm2\pi/3$ stand for the multiples of $2\pi/3$
mod $2\pi$. There are $2^4\times3=\alert{48}$ different PSG's in
this class (\Rmnum1)(B).\\

(\Rmnum{2})~~~$g_T(s)=\imth \tau^3$:

(\rmnum{1})~~~$\eta_{C_6\bs T}=1$:

(A)~~~$g_\bss(s)=\lambda_{C_6}^sg_{C_6}^{-1}(1-s)$:

in this case $\lambda_{C6}=\eta_\bss\eta_{C_6}\eta_{12}$, so we have
\begin{eqnarray}
&g_{C_6}(0)=\tau^0,\\
&g_\bss(0)=g_{C_6}(1)=\eta_{C_6}\eta_{12}\tau^0,\notag\\
&g_\bss(1)=\eta_\bss\eta_{C_6}\eta_{12}\tau^0.\notag
\end{eqnarray}
there are $2^3=\alert8$ different PSG's in the class
(\Rmnum2)(\rmnum1)(A).

(B)~~~$g_\bss(s)g_{C_6}(1-s)=\imth\hat{n}_s\cdot\vec\tau$:

($\alpha$)~~~$\eta_{\bss\bs T}=1$, \ie $[g_\bss(s),\tau^3]=0$:

here we have
\begin{eqnarray}
&g_{C_6}(0)=\tau^0,\\
&g_{C_6}(1)=\eta_{C_6}\eta_{12}\tau^0,\notag\\
&g_\bss(0)=-\imth\eta_\bss\tau^3,\notag\\
&g_\bss(1)=\imth\tau^3.\notag
\end{eqnarray}

there are $2^3=\alert8$ different PSG's in the class
(\Rmnum2)(\rmnum1)(B)($\alpha$).

($\beta$)~~~$\eta_{\bss\bs T}=-1$, \ie $\{g_\bss(s),\tau^3\}=0$:

here we have
\begin{eqnarray}
&g_{C_6}(0)=\tau^0,\\
&g_{C_6}(1)=\eta_{C_6}\eta_{12}e^{\imth\psi_3\tau^3},\notag\\
&g_\bss(0)=-\imth\eta_\bss\tau^1,\notag\\
&g_\bss(1)=\imth\tau^1.\notag
\end{eqnarray}

there are $2^3\times3=\alert{24}$ different PSG's in the class
(\Rmnum2)(\rmnum1)(B)($\beta$) since $\psi_3=0,\pm2\pi/3$.

(\rmnum{2})~~~$\eta_{C_6\bs T}=-1$:

(A)~~~$g_\bss(s)=\lambda_{C_6}^sg_{C_6}^{-1}(1-s)$:

here we must have $\eta_{\bss\bs T}=-1$,
$\lambda_{C_6}=\eta_\bss\eta_{C_6}\eta_{12}$ and
\begin{eqnarray}
&g_{C_6}(0)=\imth\tau^1,\\
&g_{C_6}(1)=-\imth\eta_{C_6}\eta_{12}\tau^1,\notag\\
&g_\bss(0)=\imth\eta_{C_6}\eta_{12}\tau^1,\notag\\
&g_\bss(1)=-\imth\eta_\bss\eta_{C_6}\eta_{12}\tau^1.\notag
\end{eqnarray}

there are $2^3=\alert8$ different PSG's in the class
(\Rmnum2)(\rmnum2)(A).

(B)~~~$g_\bss(s)g_{C_6}(1-s)=\imth\hat{n}_s\cdot\vec\tau$:

($\alpha$)~~~$\eta_{\bss\bs T}=1$, \ie $[g_\bss(s),\tau^3]=0$:

here we have
\begin{eqnarray}
&g_{C_6}(0)=\imth\tau^1,\\
&g_{C_6}(1)=-\imth\eta_{C_6}\eta_{12}\tau^1e^{\imth\psi_3\tau^3},\notag\\
&g_\bss(0)=\tau^0,\notag\\
&g_\bss(1)=\eta_\bss\tau^0.\notag
\end{eqnarray}

there are $2^3\times3=\alert{24}$ different PSG's in the class
(\Rmnum2)(\rmnum2)(B)($\alpha$) since $\psi_3=0,\pm2\pi/3$.

($\beta$)~~~$\eta_{\bss\bs T}=-1$, \ie $\{g_\bss(s),\tau^3\}=0$:

here we have
\begin{eqnarray}
&g_{C_6}(0)=\imth\tau^1,\\
&g_{C_6}(1)=-\imth\eta_{C_6}\eta_{12}\tau^1e^{\imth\psi_3\tau^3},\notag\\
&g_\bss(0)=\imth\tau^1,\notag\\
&g_\bss(1)=-\imth\eta_\bss\tau^1.\notag
\end{eqnarray}

there are $2^3\times3=\alert{24}$ different PSG's in the class
(\Rmnum2)(\rmnum2)(B)($\beta$) since $\psi_3=0,\pm2\pi/3$.\\

To summarize, above are the $\alert{160}$ different (algebraic)
PSG's with $IGG=\{\pm\tau^0\}$ on a honeycomb lattice. They
represent different $Z_2$ spin liquid states on a honeycomb lattice,
which possess all the symmetries of the honeycomb lattice generated
by $\{\bs T,T_1,T_2,\bss,C_6\}$. We also want to emphasize that any
solution to the set of equation (\ref{T^2})-(\ref{T-sig}) may look
different, but it will be gauge equivalent to one of these 160
PSG's.

On the other hand, such a (algebraic) PSG really corresponds to a
spin liquid if and only if it can be realized by a mean-field ansatz
$\{u_{ij}\}$ on a honeycomb lattice\cite{PhysRevB.65.165113}. In fact, not all of
these algebraic PSGs can be realized by an ansatz. After the
time-reveral transformation, the mean field amplitude changes
sign\cite{PhysRevB.65.165113}: $\boldsymbol T(u_{ij})=-u_{ij}$. Gauge
transformation $G_{\boldsymbol T}$ must change the sign again:
\begin{align}
 -u_{ij}=G_{\boldsymbol T}(i)u_{ij}G_{\boldsymbol T}(j)^{\dagger}
\end{align}
If in an algebraic PSG, $G_{\boldsymbol T}(i)=\tau^0$ independent of
site, $u_{ij}$ must vanish.

Clearly at least $32$ algebraic PSG's among the total $160$ types
cannot be realized by any mean-field ansatz $\{u_{ij}\}$. These are
the PSG's with $G_{\bs T}(i)=g_{\bs T}(s)=\tau^0$ in the class
(\Rmnum1)(\Rmnum1)(A)$\&$(B). Since under time reversion $\bs T$ we
require $-u_{ij}=G_{\bs T}(i)u_{ij}G^\dagger_{\bs T}(j)=u_{ij}$,
this leads to the vanishing of all bonds $\{u_{ij}\equiv0\}$ in the
mean-field ansatz. Therefore, there are at the most $\alert{128}$
possible $Z_2$ spin liquids that can be realized by a mean-field
ansatz on a honeycomb lattice.

\section{Classification of $Z_2$ projective symmetry groups around u-RVB
ansatz}\label{app:24Z2_uRVB}

In this section we focus on those $Z_2$ spin liquids near the u-RVB
state, which is discussed in section \ref{Z2_PSG}. These $Z_2$ spin
liquids are plausibly connected to a semimetal through a continuous
phase transition. The u-RVB state is realized by the following
ansatz:
\begin{eqnarray}
u_{ij}=(-1)^{s_i}\imth\chi\tau^0
\end{eqnarray}
its mean-field bond is only nonzero between nearest neighbors
$<ij>$, which have different sublattice indices $s_i=1-s_j$. By
definition, its PSG has the following form:
\begin{eqnarray}
&G_{T_1}(x_1,x_2,s)=g_1,\label{psg_uRVB}\\
&G_{T_2}(x_1,x_2,s)=g_2,\notag\\
&G_{\boldsymbol T}(x_1,x_2,s)=(-1)^sg_{\boldsymbol T},\notag\\
&G_{C_6}(x_1,x_2,s)=(-1)^sg_{C_6},\notag\\
&G_{\boldsymbol\sigma}(x_1,x_2,s)=(-1)^sg_{\boldsymbol\sigma}.\notag
\end{eqnarray}
where $g_1,g_2,g_{\bs T},g_{C_6},g_\bss\in SU(2)$. To find out those
$Z_2$ spin liquids around such a u-RVB state, we need to trace those
PSG's with $IGG=\{\pm\tau^0\}$ that looks like (\ref{psg_uRVB}).
Consistent conditions (\ref{T^2})-(\ref{T-sig}) now corresponds to
constraints on the $SU(2)$ matrices $\{g_1,g_2,g_{\bs
T},g_{C_6},g_\bss\}$:
\begin{align}
g_1^{-1}g_{2}^{-1}g_1g_2&=\xi_{12}\tau^0,&g_{\boldsymbol T}^2&=\xi_{\boldsymbol T} \tau^0,\notag\\
g_1^{-1}g_{\boldsymbol T}^{-1}g_1g_{\boldsymbol T}&=\xi_{1\boldsymbol T}\tau^0,&g_2^{-1}g_{\boldsymbol T}^{-1}g_2g_{\boldsymbol T}&=\xi_{2\boldsymbol T}\tau^0,\notag\\
g_2^{-1}g_{C_6}g_1g_{C_6}^{-1}&=\xi_{C_6 1}\tau^0,&g_1^{-1}g_{C_6}g_1g_2^{-1}g_{C_6}^{-1}&=\xi_{C_6 2}\tau^0,\notag\\
g_{\boldsymbol T}^{-1}g_{C_6}^{-1}g_{\boldsymbol T}g_{C_6}&=\xi_{C_6\boldsymbol T}\tau^0,&g_{C_6}^6&=\xi_{C_6}\tau^0,\notag\\
g_1^{-1}g_{\boldsymbol \sigma}g_1g_{\boldsymbol \sigma}^{-1}&=\xi_{\boldsymbol \sigma 1}\tau^0,&g_2^{-1}g_{\boldsymbol\sigma}g_1g_2^{-1}g_{\boldsymbol\sigma}^{-1}&=\xi_{\boldsymbol\sigma 2}\tau^0,\notag\\
g_{\boldsymbol\sigma}g_{C_6}g_{\boldsymbol\sigma}g_{C_6}&=\xi_{\boldsymbol\sigma C_6}\tau^0,&g_{\boldsymbol T}^{-1}g_{\boldsymbol\sigma}^{-1}g_{\boldsymbol T}g_{\boldsymbol\sigma}&=\xi_{\boldsymbol\sigma \boldsymbol T}\tau^0,\notag\\
g_{\boldsymbol\sigma}^2&=\xi_{\boldsymbol\sigma}\tau^0.\label{g_urvb_constraint}
\end{align}
where all $\xi$'s take value of $\pm1$. Again, as discussed in
appendix \ref{app:160Z2} we can always make
$\xi_{C_61}=\xi_{C_62}=1$ by choosing a proper gauge. After solving
eqs. (\ref{g_urvb_constraint}), we find out there are $\alert{24}$
gauge inequivalent solutions in total, as summarized in TABLE
\ref{tab:z2_urvb}. In other words, there are $24$ different $Z_2$
spin liquid around the u-RVB state, suggesting that they are
promising candidates of the spin liquid connected to a semimetal on
honeycomb lattice through a continuous phase transition.

\begin{table}[tb]
\begin{tabular}{|c||c|c|c|c|c|}
\hline $\#$ & $g_{\bs T}$ & $g_\bss$ & $g_{C_6}$& $g_1$&$g_2$\\
\hline 1&$\tau^0$&$\tau^0$&$\tau^0$&$\tau^0$&$\tau^0$\\
\hline 2&$\tau^0$&$\tau^0$&$i\tau^3$&$\tau^0$&$\tau^0$\\
\hline 3&$\tau^0$&$\tau^0$&$i\tau^3$&$e^{\imth2\pi/3\tau^1}$&$e^{-\imth2\pi/3\tau^1}$\\
\hline 4&$\tau^0$&$\imth\tau^3$&$\imth\tau^3$&$\tau^0$&$\tau^0$\\
\hline 5&$\tau^0$&$\imth\tau^3$&$\imth\tau^3$&$\tau^0$&$\tau^0$\\
\hline 6&$\tau^0$&$\imth\tau^3$&$\imth\tau^1$&$\tau^0$&$\tau^0$\\
\hline 7&$\tau^0$&$\imth\tau^3$&$e^{\imth\pi/6\tau^1}$&$\tau^0$&$\tau^0$\\
\hline 8&$\tau^0$&$\imth\tau^3$&$e^{\imth\pi/3\tau^1}$&$\tau^0$&$\tau^0$\\
\hline 9&$\tau^0$&$\imth\tau^3$&$\imth\tau^1$&$e^{\imth2\pi/3\tau^3}$&$e^{-\imth2\pi/3\tau^3}$\\
\hline 10&$\tau^0$&$\imth\tau^3$&$e^{\imth2\pi/3\tau^1}$&$\imth(\frac{\tau^1}{\sqrt3}-\sqrt{\frac23}\tau^2)$&$\imth(\frac{\tau^3}{\sqrt2}-\frac{\tau^2}{\sqrt6}-\frac{\tau^1}{\sqrt3})$\\
\hline 11&$\imth\tau^3$&$\tau^0$&$\tau^0$&$\tau^0$&$\tau^0$\\
\hline 12&$\imth\tau^3$&$\tau^0$&$\imth\tau^3$&$\tau^0$&$\tau^0$\\
\hline 13&$\imth\tau^3$&$\tau^0$&$\imth\tau^1$&$\tau^0$&$\tau^0$\\
\hline 14&$\imth\tau^3$&$\tau^0$&$\imth\tau^1$&$e^{\imth2\pi/3\tau^3}$&$e^{-\imth2\pi/3\tau^3}$\\
\hline 15&$\imth\tau^3$&$\imth\tau^3$&$\tau^0$&$\tau^0$&$\tau^0$\\
\hline 16&$\imth\tau^3$&$\imth\tau^3$&$\imth\tau^3$&$\tau^0$&$\tau^0$\\
\hline 17&$\imth\tau^3$&$\imth\tau^3$&$\imth\tau^1$&$\tau^0$&$\tau^0$\\
\hline 18&$\imth\tau^3$&$\imth\tau^3$&$\imth\tau^1$&$e^{\imth2\pi/3\tau^3}$&$e^{-\imth2\pi/3\tau^3}$\\
\hline 19&$\imth\tau^3$&$\imth\tau^1$&$\imth\tau^1$&$\tau^0$&$\tau^0$\\
\hline 20&$\imth\tau^3$&$\imth\tau^1$&$\imth\tau^2$&$\tau^0$&$\tau^0$\\
\hline 21&$\imth\tau^3$&$\imth\tau^1$&$\tau^0$&$\tau^0$&$\tau^0$\\
\hline 22&$\imth\tau^3$&$\imth\tau^1$&$\imth\tau^3$&$\tau^0$&$\tau^0$\\
\hline 23&$\imth\tau^3$&$\imth\tau^1$&$e^{\imth\pi/6\tau^3}$&$\tau^0$&$\tau^0$\\
\hline 24&$\imth\tau^3$&$\imth\tau^1$&$e^{\imth\pi/3\tau^3}$&$\tau^0$&$\tau^0$\\
\hline
\end{tabular}
\caption{\label{tab:z2_urvb}A summary of all 24 different PSG's with
$IGG=\{\pm\tau^0\}$ around the u-RVB ansatz. They correspond to 24
different $Z_2$ spin liquids near the u-RVB state.}
\end{table}

\section{Consistent conditions on the mean-field ansatz $\{u_{ij}\}$ on a honeycomb
lattice}\label{app:c.c.u_ij}

In this section we derive the consistent conditions on an arbitrary
mean-field bond $u_{ij}$, which realizes a spin liquid with a
certain PSG on a honeycomb lattice. The basic idea is to find all
possible symmetry group elements that transform the two lattice
sites $\{i,j\}$ into itself $\{i,j\}$ or into each other $\{j,i\}$,
so that the corresponding PSG elements must transform mean-field
bond $u_{ij}$ into itself $u_{ij}$ or its Hermitian conjugate
$u_{ij}^\dagger=u_{ji}$.

As a special case, the identity element $1$ always transform a bond
into itself: correspondingly in $PSG$ the $IGG$ elements (\eg
$\tau^0$ for a $Z_2$ ansatz) always transform any bond $u_{ij}$ into
itself. This is nothing but the definition of invariant gauge group
(IGG).

Now we need to look at nontrivial symmetry group elements which
transform two lattice sites (connected by the bond) into itself or
into each other. Without loss of generality, we consider the
following bond
\begin{eqnarray}
\langle x_1,x_2,s\rangle\equiv u_{(x_1,x_2,s)(0,0,0)}
\end{eqnarray}

\subsection{Regarding time reversal $\bs T$}

Any element of the symmetry group can be written as
\begin{eqnarray}
U=\bs{T}^{\nu_{\bs{T}}}\cdot T_1^{\nu_{T_1}}\cdot
T_2^{\nu_{T_2}}\cdot\bs{\sigma}^{\nu_{\bs{\sigma}}}\cdot
C_6^{\nu_{C_6}}
\end{eqnarray}
First we study the consistent conditions from time reversal
transformation $\bs T$ and then turn to other group elements.

Notice that time reversal $\bs T$ doesn't change anything except the
sign of bond:
\begin{eqnarray}
G_{\bs T}(i)u_{ij}[G_{\bs T}(j)]^\dagger=-u_{ij}
\end{eqnarray}
so this bond must satisfy the following constraint:
\begin{eqnarray}
&G_{\bs T}(x_1,x_2,s)\langle x_1,x_2,s\rangle\\
&=-\langle x_1,x_2,s\rangle G_{\bs T}(0,0,0)\notag
\end{eqnarray}

\subsection{Conditions on a bond within the same sublattice: $s=0$}

First we study $s=0$ case, \ie a bond within the same sublattice.
Since both mirror reflection $\bss$ and $\pi/3$ rotation $C_6$ will
change the sublattice index $s$ while the translations $T_1,T_2$
don't, we must have an even number of reflection and rotation, \ie
$\nu_\bss+\nu_{C_6}=0\mod2$ to transform the bond to itself (or its
Hermitian conjugate).

>From (\ref{sg_transf}) it's easy to check the 5 nontrivial elements
consisting of $\{\bss,~C_6\}$:
\begin{eqnarray}
&C_6^2(x_1,x_2,0)=(1-x_1-x_2,x_1,0),\notag\\
&C_6^{-2}(x_1,x_2,0)=(x_2,1-x_1-x_2,0),\notag\\
&\bss C_6(x_1,x_2,0)=(x_1,1-x_1-y_1,0),\notag\\
&\bss C_6^3(x_1,x_2,0)=(1-x_1-x_2,x_2,0),\notag\\
&\bss C_6^{-1}(x_1,x_2,0)=(x_2,x_1,0).
\end{eqnarray}
In order that the bond goes back after some translations, it's
straightforward to find out all the consistent conditions on such a
bond:
\begin{eqnarray}\label{c.c._s=0}
&T_2^{-1}\bss C_6:~~\langle -2x,x,0\rangle\rightarrow\langle
-2x,x,0\rangle\\
&T_2^{x-1}\bss C_6:~~\langle 0,x,0\rangle\rightarrow\langle
0,x,0\rangle^\dagger\notag\\
&T_1^{-1}\bss C_6^3:~~\langle x,-2x,0\rangle\rightarrow\langle
x,-2x,0\rangle\notag\\
&T_1^{x-1}\bss C_6^3:~~\langle x,0,0\rangle\rightarrow\langle
x,0,0\rangle^\dagger\notag\\
&\bss C_6^{-1}:~~\langle x,x,0\rangle\rightarrow\langle
x,x,0\rangle\notag\\
&T_1^{x}T_2^{-x}\bss C_6^{-1}:~~\langle
x,-x,0\rangle\rightarrow\langle x,-x,0\rangle^\dagger\notag
\end{eqnarray}
for $\forall~~x\in\mathbb{Z}$.

\subsection{Conditions on a bond connecting different sublattices: $s=1$}

In the $s=1$ case, such a bond connects different sublattices. So
only an even number of reflection and rotation, \ie
$\nu_\bss+\nu_{C_6}=0\mod2$ might transform the bond to itself,
while an odd number of reflection and rotation, \ie
$\nu_\bss+\nu_{C_6}=1\mod2$ can transform the bond $\langle
x_1,x_2,1\rangle$ into its Hermitian conjugate $\langle x_1,x_2,1
\rangle^\dagger$.

It's straightforward to obtain the following conditions on the bond
$\langle x_1,x_2,1\rangle\equiv u_{(x_1,x_2,1)(0,0,0)}$:
\begin{eqnarray}\label{c.c._s=1}
&\bss:~~\langle -2x,x,1\rangle\rightarrow\langle
-2x,x,1\rangle^\dagger\\
&\bss C_6^{-1}:~~\langle x+1,x,1\rangle\rightarrow\langle
x+1,x,1\rangle\notag\\
&T_1^{-2x-2}T_2^{x+1}\bss C_6^{-2}:~~\langle
-2x-1,x,1\rangle\rightarrow\langle
-2x-1,x,1\rangle^\dagger\notag\\
&T_1^{x_1-1}T_2^{x_2}C_6^3:~~\langle
x_1,x_2,1\rangle\rightarrow\langle
x_1,x_2,1\rangle^\dagger\notag\\
&T_1^{-1}\bss C_6^3:~~\langle x,-2x,1\rangle\rightarrow\langle
x,-2x,1\rangle\notag\\
&T_1^{x-1}T_2^{x-1}\bss C_6^2:~~\langle
x+1,x,1\rangle\rightarrow\langle x+1,x,1\rangle^\dagger\notag\\
&T_2^{-1}\bss C_6:~~\langle -2x-1,x,1\rangle\rightarrow\langle
-2x-1,x,1\rangle\notag
\end{eqnarray}
for $\forall~~x,x_1,x_2\in\mathbb{Z}$.

\subsection{An example: mean-field ansatz $\{u_{ij}\}$ of $Z_2$ spin liquids near u-RVB state}

To demonstrate the above consistent conditions, let's take a look at
how they determine the mean-field ansatz $\{u_{ij}\}$ of any $Z_2$
spin liquid near u-RVB state, with PSG generators (\ref{psg_uRVB}).

Considering time reversion $\bs T$ we immediately have
\begin{eqnarray}\label{c.c._T}
g_{\bs T}\langle x_1,x_2,s\rangle=-(-1)^s\langle x_1,x_2,s\rangle
g_{\bs T}
\end{eqnarray}\\
In other words, the bond connecting two sites belonging to the same
(different) sublattice(s) anti-commutes(commutes) with $g_{\bs T}$.

For the nearest neighbor (n.n.) bond
$u_\alpha\equiv\langle0,0,1\rangle$ we have $x_1=x_2=0,s=1$.
Conditions (\ref{c.c._s=1}) and (\ref{c.c._T}) immediately lead to
\begin{eqnarray}
&[u_\alpha,g_{\bs T}]=0\\
&g_\bss u_\alpha=-u_\alpha^\dagger g_\bss\notag\\
&g_1^{-1}g_{C_6}^3u_\alpha=-u_\alpha^\dagger g_1^{-1}g_{C_6}^3\notag
\end{eqnarray}

For 2nd n.n. bond $u_\beta\equiv\langle0,1,0\rangle$ we have
$x_1=0=s,x_2=1$ and (\ref{c.c._s=0}), (\ref{c.c._T}) lead to
\begin{eqnarray}
&\{u_\beta,g_{\bs T}\}=0\\
&g_\bss g_{C_6}u_\beta=u_\beta^\dagger g_\bss g_{C_6}\notag
\end{eqnarray}

For 3rd n.n. bond $u_\gamma\equiv\langle1,0,1\rangle$ we have
$x_2=0,x_1=s=1$. Conditions (\ref{c.c._s=1}) and (\ref{c.c._T}) lead
to
\begin{eqnarray}
&[u_\gamma,g_{\bs T}]=0\\
&g_{C_6}^3 u_\gamma=-u_\gamma^\dagger g_{C_6}^3\notag\\
&g_\bss g_{C_6}^{-1}u_\gamma=u_\gamma g_\bss g_{C_6}^{-1}\notag
\end{eqnarray}

Constraints on further neighbors: \eg 4th n.n.
$\langle0,1,1\rangle$, 5th n.n. $\langle1,1,0\rangle$ and 6th n.n.
$\langle2,0,0\rangle$ can be similarly obtained.

\section{A search of gapped spin liquids near the u-RVB
state}\label{app:mass}

In appendix \ref{app:24Z2_uRVB} we showed that there are at most 24
$Z_2$ spin liquids around the u-RVB state, which are likely to
connect with a semimetal through a continuous phase transition. In
this section we search for those states with spectral gaps among the
24 spin liquid ansatz. In the end we find out most of the 24 states
are gapless. More specifically, they cannot open up a mass gap
through any perturbation around the u-RVB state, which has two
graphenelike Dirac cones in the 1st Brillouin zone. It turns out
that only 4 of them, \ie $\#16,~\#17,~\#19$ and $\#22$ in TABLE
\ref{tab:z2_urvb}, are gapped spin liquids near the u-RVB state.

\subsection{Symmetry-allowed masses in a graphenelike u-RVB state}

We start from the low-energy effective Hamiltonian of the u-RVB
state, which is described by a massless 8-component Dirac equation.
These 8 components contain 2 spin indices (labeled by Pauli matrices
$\{\tau^i\}$), 2 sublattice indices (labeled by Pauli matrices
$\{\mu^i\}$) and 2 valley indices (labeled by Pauli matrices
$\{\nu^i\}$). Just like graphene, the two valleys are located at
$\bf{K}$ and $\bf{K^\prime}$, \ie the vertices in the
honeycomb-shaped 1st Brillouin zone. Following the convention shown
in FIG. \ref{fig:honeycomb}, the momentum of these two cones are
${\bf{K}}={4\pi\over3}\vec b_1+{2\pi\over3}\vec b_2$ and
${\bf{K^\prime}}={2\pi\over3}\vec b_1+{4\pi\over3}\vec b_2$
respectively, where $\{\vec b_1=(\sqrt3,-1)/\sqrt3a,~\vec
b_2=(0,2)/\sqrt3a\}$ are the reciprocal lattice vectors
corresponding to lattice vectors $\{\vec a_1=(a,0),~\vec
a_2=(1,\sqrt3)a/2\}$.

Expanding the mean-field Hamiltonian of a u-RVB state with
$u_\alpha=\imth\tau^0$ (here ${\bf
k}=\frac{2}{\sqrt3a}(k_x,k_y)=k_1\vec b_1+k_2\vec b_2$)
\begin{eqnarray}
&H_{uRVB}=\imth(\psi^\dagger_{{\bf k},A},\psi^\dagger_{{\bf k},B})\cdot\notag\\
&\begin{bmatrix}0&-\tau^0(1+e^{-\imth k_2}+e^{\imth(k_1-k_2)})\\
\tau^0(1+e^{\imth
k_2}+e^{\imth(k_2-k_1)})&0\end{bmatrix}\notag\\
&\cdot\begin{pmatrix}\psi_{{\bf k},A}\\
\psi_{{\bf k},B}\end{pmatrix}\notag
\end{eqnarray}
around $\bf K$ and $\bf K^\prime$ we immediately obtain the Dirac
equations
\begin{eqnarray}
&H_{\bf K}=(\psi^\dagger_{{\bf k},A},\psi^\dagger_{{\bf k},B})\begin{bmatrix}0&\tau^0(k_y+\imth k_x)\\ \tau^0(k_y-\imth k_x)&0\end{bmatrix}\begin{pmatrix}\psi_{{\bf k},A}\\
\psi_{{\bf k},B}\end{pmatrix}\notag\\
&H_{\bf K^\prime}=(\psi^\dagger_{{\bf k^\prime},A},\psi^\dagger_{{\bf k^\prime},B})\begin{bmatrix}0&\tau^0(k_y^\prime-\imth k_x^\prime)\\ \tau^0(k_y^\prime+\imth k_x^\prime)&0\end{bmatrix}\begin{pmatrix}\psi_{{\bf k^\prime},A}\\
\psi_{{\bf k^\prime},B}\end{pmatrix}\notag
\end{eqnarray}
Defining the following 8-component spinor:
\begin{eqnarray}
\Psi^T\equiv(\psi_{{\bf k},A}^T,\psi_{{\bf k},B}^T,\psi_{{\bf
k^\prime},B}^T,\psi_{{\bf k^\prime},A}^T)
\end{eqnarray}
we can write the above effective Hamiltonian of u-RVB state as
\begin{eqnarray}
H=\Psi^\dagger\mu^3(\mu^2\partial_x+\mu^1\partial_y)\otimes\tau^0\otimes\nu^0\Psi
\end{eqnarray}

Therefore only those mass terms
$M=\mu^3\otimes\tau^a\otimes\nu^b,~~a,b=0,1,2,3$ satisfy that
$\{H,\Psi^\dagger M\Psi\}=0$ so that a mass gap can be opened in the
Dirac spectrum. In the following we study how the mass term changes
under the action of symmetry transformation such as spin rotations,
time reversal $\bs T$ and translations $T_1,T_2$ \etc The physical
symmetry of a spin liquid state realized by mean-field ansatz only
allow those masses that are invariant under the corresponding PSG.
If a PSG already rules out all possible mass terms
$M=\mu^3\otimes\tau^a\otimes\nu^b,~~a,b=0,1,2,3$, we conclude the
corresponding spin liquid realized by mean-field ansatz is gapless.

First we work out the transformation rules of Dirac spinor $\Psi$
and $M$ under a PSG. We focus on the 24 PSG's near the u-RVB state
with the form (\ref{psg_uRVB}) as summarized in TABLE
\ref{tab:z2_urvb}.

\subsubsection{Spin rotations}% along $\hat{y}$-,~$\hat{z}$-axis}

It's straightforward to show that a spin rotation along
$\hat{z}$-axis by $2\theta$ angle is realized by
\begin{eqnarray}
\Psi\rightarrow e^{\imth \theta}\Psi
\end{eqnarray}
while a spin rotation along $\hat{y}$-axis by $\pi$ angle is
realized by
\begin{eqnarray}
\Psi\rightarrow \imth \tau^2\mu^1\nu^1\Psi^\ast
\end{eqnarray}

Apparently $S_z$ rotations leave the mass term invariant, while
under $S_y$ rotations by $\pi$ the mass term transforms in the
following way
\begin{eqnarray}
M\rightarrow-\mu^1\otimes\nu^1\otimes\tau^2
M^T\tau^2\otimes\mu^1\otimes\nu^1
\end{eqnarray}

Since the mass term is invariant under spin rotations, its allowed
form as seen from the above constraint can only be
\begin{eqnarray}
M_A^{(a)}=\mu^3\otimes\nu^3\otimes\tau^a,~~~a=1,2,3
\end{eqnarray}
or
\begin{eqnarray}
M_B^{(b)}=\mu^3\otimes\nu^b\otimes\tau^0,~~~b=0,1,2
\end{eqnarray}

\subsubsection{Time reversal $\bs T$}

Since a mean-field bond $u_{ij}$ becomes $-(-1)^{s_i}g_T
u_{ij}g^\dagger_T(-1)^{s_j}$ under the time reversal transformation
in a PSG (\ref{psg_uRVB}), clearly $\bs T$ is realized by
\begin{eqnarray}
&\Psi\rightarrow g_T^\dagger\otimes\mu^3\otimes\nu^3\Psi\notag\\
&M\rightarrow-M
\end{eqnarray}
so the mass term is invariant under time reversal $\bs T$ if
\begin{eqnarray}\label{mass:T}
M=-g_T\otimes\mu^3\otimes\nu^3M g_T^\dagger\otimes\mu^3\otimes\nu^3
\end{eqnarray}

10 spin liquids near the u-RVB state, \ie \#1-\#10 in TABLE
\ref{tab:z2_urvb} has $g_T=\tau^0$. In these cases, mass terms
$M_A^{(a)},~a=1,2,3$ will violate transformation rule
(\ref{mass:T}), and the only allowed masses are
$M_B^{(1)},~M_B^{(2)}$.

The other 14 spin liquids around u-RVB state (\#11-\#24 in TABLE
\ref{tab:z2_urvb}) are characterized by $g_T=\imth\tau^3$. In this
case the allowed masses are $M_B^{(1)},~M_B^{(2)}$ and
$M_A^{(1)},~M_A^{(2)}$.

\subsubsection{Translations $T_1,~T_2$}

Under translations $T_1,T_2$ in a PSG (\ref{psg_uRVB}) the
8-component spinor changes as
\begin{eqnarray}
T_1:~~~&\Psi\rightarrow e^{-\imth\frac{2\pi}3\nu^3}\otimes g_1^\dagger\Psi,\notag\\
T_2:~~~&\Psi\rightarrow e^{\imth\frac{2\pi}3\nu^3}\otimes
g_2^\dagger\Psi.
\end{eqnarray}
since ${\bf K}\cdot\vec a_{1,2}=\mp\frac{2\pi}3$ and ${\bf
K^\prime}\cdot\vec a_{1,2}=\pm\frac{2\pi}3$. In order for the mass
term to be invariant
\begin{eqnarray}
&M=e^{\imth\frac{2\pi}3\nu^3}\otimes
g_1Me^{-\imth\frac{2\pi}3\nu^3}\otimes g_1^\dagger\notag\\
&=e^{-\imth\frac{2\pi}3\nu^3}\otimes
g_2Me^{\imth\frac{2\pi}3\nu^3}\otimes g_2^\dagger\label{mss:T1T2}
\end{eqnarray}
the symmetry-allowed masses can only be:

$M_B^{(0)}$ and $M_A^{(a)},~a=1,2,3$ if $g_1=g_2=\tau^0$;

$M_B^{(0)}$ and $M_A^{(3)}$ if $g_1=g_2^{-1}=e^{\imth2\pi/3\tau^3}$;

$M_B^{(0)}$ for the special case \#10 in TABLE \ref{tab:z2_urvb}.\\

Combining conditions (\ref{mass:T}) and (\ref{mss:T1T2}) we can see
that $\{M_B^{(b)},~b=0,1,2\}$ are not allowed by symmetry in any of
the 24 spin liquids near u-RVB state. In the following study we will
focus on masses $M_A^{(a)},~a=1,2,3$.

\subsubsection{Reflection $\bss$}

Similar to time reversal $\bs T$, under reflection along
$\hat{x}$-axis the spinor transforms as
\begin{eqnarray}
\Psi\rightarrow \mu^1\cdot
g_\bss^\dagger\otimes\mu^3\otimes\nu^3\Psi=-\imth
g_\bss^\dagger\otimes\mu^2\otimes\nu^3\Psi
\end{eqnarray}
The mass term is invariant under reflection $\bss$ if
\begin{eqnarray}
M=g_\bss\otimes\mu^2\otimes\nu^3Mg_\bss^\dagger\otimes\mu^2\otimes\nu^3\label{mass:sig}
\end{eqnarray}

The symmetry-allowed masses are:

none if $g_\bss=\tau^0$;

$M_A^{(a)},~a\neq b$ if $g_\bss=\imth\tau^b$.

\subsubsection{$\pi/3$ rotation $C_6$}

Under $C_6$, \ie a rotation by $\pi/3$ the spinor transforms as
\begin{eqnarray}
\Psi\rightarrow g_{C_6}^\dagger\otimes
e^{\imth\frac{5\pi}6\mu^3}\otimes(\frac{\sqrt3}2\nu^1-\frac12\nu^2)\Psi
\end{eqnarray}
The mass term is invariant under reflection $\bss$ if
\begin{eqnarray}
&M=g_{C_6}\otimes
e^{-\imth\frac{5\pi}6\mu^3}\otimes(\frac{\sqrt3}2\nu^1-\frac12\nu^2)\cdot M\notag\\
&\cdot g_{C_6}^\dagger\otimes
e^{\imth\frac{5\pi}6\mu^3}\otimes(\frac{\sqrt3}2\nu^1-\frac12\nu^2)\label{mass:c6}
\end{eqnarray}

The symmetry-allowed masses are:

none if $g_{C_6}=\tau^0,~e^{\imth\theta\tau^{1,3}}$ with
$\theta\neq0\mod\pi/2$;

$M_A^{(a)},~a\neq b$ if $g_{C_6}=\imth\tau^b$.

\subsection{Realizing the 4 gapped $Z_2$ spin liquids near the u-RVB
state}\label{app:mass:4_z2}

Among all 24 spin liquids near the u-RVB states, it turns out that
there are no symmetry-allowed masses for 20 of them. In other words,
these 20 spin liquids cannot open up a mass gap through a
perturbation around the u-RVB state. Only the following 4 spin
liquids near the u-RVB state can obtain an energy gap in the
spectrum through adding a symmetry-allowed mass term:

\#16 with two symmetry-allowed masses
$M_A^{(1,2)}=\mu^3\otimes\nu^{3}\otimes\tau^{1,2}$;

\#17 with one symmetry-allowed mass
$M_A^{(2)}=\mu^3\otimes\nu^{3}\otimes\tau^2$;

\#19 with one symmetry-allowed mass
$M_A^{(2)}=\mu^3\otimes\nu^{3}\otimes\tau^2$;

\#22 with one symmetry-allowed mass
$M_A^{(2)}=\mu^3\otimes\nu^{3}\otimes\tau^2$.

In fact, as summarized in TABLE \ref{tab:4z2_ansatz}, all these 4
gapped spin liquids can be realized by mean-field ansatz
$\{u_{ij}\}$, which satisfies consistent conditions from the
corresponding PSG as discussed in appendix \ref{app:c.c.u_ij}. In
the following we describe the mean-field ansatz for these 4 gapped
$Z_2$ spin liquids. In the end only one gapped $Z_2$ spin liquid,
\ie \#19 can be realized by a mean-field ansatz up to 3rd n.n.
bonds.

\begin{table}[tb]
\begin{tabular}{|c||c|c|c|c|c|c|}
\hline $\#$ & $u_\alpha$ & $u_\beta$ & $u_\gamma$& $u_\delta$&$u_\varepsilon$&9th n.n. $\langle1,2,0\rangle$\\
\hline 16&$\imth\tau^0$&$\{\tau^1,\tau^2\}$&$\imth\tau^0$&$\imth\tau^0$&$\{\tau^1,\tau^2\}$&$\cdots$\\
\hline 17&$\imth\tau^0$&$\tau^2$&$\imth\tau^0$&$\{\imth\tau^0,\tau^3\}$&$\cdots$&$\cdots$\\
\hline 19&$\{\imth\tau^0,\tau^3\}$&$\{\tau^1,\tau^2\}$&$\cdots$&$\cdots$&$\cdots$&$\cdots$\\
\hline 22&$\imth\tau^0$&$\tau^2$&$\imth\tau^0$&$\imth\tau^0$&$\tau^2$&$\{\tau^1,\tau^2\}$\\
\hline
\end{tabular}
\caption{\label{tab:4z2_ansatz}Symmetry-allowed mean-field ansatz of
the 4 possible gapped spin liquids near the u-RVB state. We follow
the notations for mean-field bonds in appendix \ref{app:c.c.u_ij}.
We only summarize the mean-field bonds that are necessary to realize
a gapped $Z_2$ spin liquid. Ellipsis represents those longer-range
mean-field bonds unnecessary for a $Z_2$ spin liquid, which are not
listed in this table. Up to 3rd n.n. mean-field bonds
$\{u_\alpha,u_\beta,u_\gamma\}$, only one $Z_2$ spin liquid, \ie
\#19 can be realized on a honeycomb lattice.}
\end{table}

\subsubsection{$Z_2$ spin liquid \#16:~up to 5th n.n. bonds needed}

The mean-field ansatz $\{u_{ij}\}$ for $Z_2$ spin liquid \#16 is
summarized in TABLE \ref{tab:4z2_ansatz}, up to 5th n.n. bonds. The
corresponding spin liquid has a $Z_2$ gauge structure, if and only
if $[u_\beta,u_\varepsilon]\neq0$, so that the IGG of this
mean-field ansatz is a $Z_2$ group $\{\pm\tau^0\}$.

It's straightforward to check that 2nd n.n. bond
$u_\beta=\beta_1\tau^1+\beta_2\tau^2$ open up a mass gap
$M\sim\mu^3\otimes\nu^3\otimes(\beta_1\tau^1+\beta_2\tau^2)=\beta_1M_A^{(1)}+\beta_2M_A^{(2)}$.

\subsubsection{$Z_2$ spin liquid \#17:~up to 4th n.n. bonds needed}

The mean-field ansatz $\{u_{ij}\}$ for $Z_2$ spin liquid \#17 is
summarized in TABLE \ref{tab:4z2_ansatz}, up to 4th n.n. bonds. The
corresponding spin liquid has a $Z_2$ gauge structure, if and only
if $[u_\beta,u_\delta]\neq0$, so that the IGG of this mean-field
ansatz is a $Z_2$ group $\{\pm\tau^0\}$.

It's straightforward to check that 2nd n.n. bond
$u_\beta=\beta\tau^2$ open up a mass gap
$M\sim\beta\mu^3\otimes\nu^3\otimes\tau^2=\beta M_A^{(2)}$.

\subsubsection{$Z_2$ spin liquid \#19:~up to 2nd n.n. bonds needed}

The mean-field ansatz $\{u_{ij}\}$ for $Z_2$ spin liquid \#17 is
summarized in TABLE \ref{tab:4z2_ansatz}, up to 2nd n.n. bonds. The
corresponding spin liquid has a $Z_2$ gauge structure, if and only
if $u_\beta=\beta_1\tau^1+\beta_2\tau^2$ with
$\beta_1,\beta_2\neq0$, so that the IGG of this mean-field ansatz is
a $Z_2$ group $\{\pm\tau^0\}$. This is the only gapped $Z_2$ spin
liquid near the u-RVB state, that can be realized in a mean-field
ansatz up to 3rd n.n. bonds.

It's straightforward to check that 2nd n.n. bond
$u_\beta=\beta_1\tau^1+\beta_2\tau^2$ open up a mass gap
$M\sim\beta_2\mu^3\otimes\nu^3\otimes\tau^2=\beta_2M_A^{(2)}$.

\subsubsection{$Z_2$ spin liquid \#22:~up to 9th n.n. bonds needed}

The mean-field ansatz $\{u_{ij}\}$ for $Z_2$ spin liquid \#17 is
summarized in TABLE \ref{tab:4z2_ansatz}, up to 9th n.n. bonds. The
corresponding spin liquid has a $Z_2$ gauge structure, if and only
if $[u_\beta,u_9]\neq0$, so that the IGG of this mean-field ansatz
is a $Z_2$ group $\{\pm\tau^0\}$. $u_9\equiv\langle1,2,0\rangle$ is
the 9th n.n. mean-field bond. In this $Z_2$ spin liquid, the
symmetry-allowed consistent mean-field bonds for 6th, 7th and 8th
n.n. are:
\begin{eqnarray}
&u_6\equiv\langle2,0,0\rangle\propto\tau^2,\notag\\
&u_7\equiv\langle2,0,1\rangle\propto\imth\tau^0,\notag\\
&u_8\equiv\langle0,2,1\rangle\propto\imth\tau^0.\notag
\end{eqnarray}

It's straightforward to check that 2nd n.n. bond
$u_\beta=\beta\tau^2$ open up a mass gap
$M\sim\beta\mu^3\otimes\nu^3\otimes\tau^2=\beta M_A^{(2)}$.

%\bibliographystyle{apsrev}
%\bibliography{/home/ranying/downloads/reference/simplifiedying}

\end{document}